\documentclass[prd,nofootinbib,floats,aps,twocolumn,tightenlines,superscriptaddress]
{revtex4-2}

\usepackage{natbib}
\usepackage{cancel}
\usepackage[skins,breakable]{tcolorbox}
\usepackage{float} 
\usepackage{graphicx}
\usepackage{mathtools}
\usepackage[normalem]{ulem}
\usepackage{hyperref}
\hypersetup{
	colorlinks=true,
	linkcolor=blue,
	filecolor=magenta,      
	urlcolor=cyan,
}
\usepackage{commath}
\usepackage{bm}
\usepackage{amsfonts,amsmath,amssymb,amsthm}
\usepackage[usenames,dvipsnames]{xcolor}
\usepackage{dsfont}
\usepackage{enumitem}
\usepackage{color}
\usepackage{tikz}
\usepackage{physics}
\usepackage{verbatim}
\usepackage{subfigure}

\usepackage{braket}

\newcommand{\ii}{\mathrm{i}}

\renewcommand*\d[2][]{%
	\mathrm{d}%
	\ifx\relax#1\relax\else
	\rule{-0.02em}{1.5ex}^{#1}\rule{0.08em}{0ex}\!
	\fi
	#2\,
}

\def\be{\begin{equation}}
\def\ee{\end{equation}}

\makeatletter 

\renewcommand\onecolumngrid{
	\do@columngrid{one}{\@ne}%
	\def\set@footnotewidth{\onecolumngrid}
	\def\footnoterule{\kern-6pt\hrule width 1.5in\kern6pt}%
}

\renewcommand\twocolumngrid{
	\def\footnoterule{
		\dimen@\skip\footins\divide\dimen@\thr@@
		\kern-\dimen@\hrule width.5in\kern\dimen@}
	\do@columngrid{mlt}{\tw@}
}%
\makeatother

\newcommand\restr[2]{{
		\left.\kern-\nulldelimiterspace 
		#1 
		\vphantom{\normal|} 
		\right|_{#2} 
}}

\definecolor{ForestGreen}{RGB}{35,120,35}

\newtheorem{theorem}{Theorem}
\newtheorem{proposition}{Proposition}
\newtheorem{lemma}{Lemma}
\newtheorem*{claim*}{Claim}

\theoremstyle{definition}

\theoremstyle{remark}

\begin{document}

\title{Partner-mode overlap as a symplectic-invariant\\ measure of correlations in Gaussian quantum field theories}
	
\author{Ivan Agullo}
\email{agullo@lsu.edu}
\affiliation{Department of Physics and Astronomy, Louisiana State University, Baton Rouge, LA 70803, USA}
\author{Eduardo Mart\'{i}n-Mart\'{i}nez}
\email{emartinmartinez@uwaterloo.ca}
\affiliation{Department of Applied Mathematics, University of Waterloo, Waterloo, Ontario, N2L 3G1, Canada}
\affiliation{Institute for Quantum Computing, University of Waterloo, Waterloo, Ontario, N2L 3G1, Canada}
\affiliation{Perimeter Institute for Theoretical Physics, Waterloo, Ontario, N2L 2Y5, Canada}

\author{Sergi Nadal-Gisbert}
\email{sergi.nadalgisbert@helsinki.fi}
\affiliation{QTF Centre of Excellence, Department of Physics,
University of Helsinki, FI-00014 Helsinki, Finland}

\author{Koji Yamaguchi}
\email{koji.yamaguchi@uwaterloo.ca}
\affiliation{Department of Informatics, Faculty of Information Science and Electrical Engineering,
Kyushu University, 744 Motooka, Nishi-ku, Fukuoka, 819-0395, Japan}

\begin{abstract}

We introduce a locally symplectic-invariant quantifier of correlations between arbitrary bosonic Gaussian modes, with particular emphasis on quantum field theory. The quantity, denoted by~$\mathcal{D}^{\mathrm{sym}}$,  admits a simple geometric interpretation as the symmetric overlap between each mode and the purification partner of the other, providing a direct geometric characterization of how correlations are distributed between modes. We derive a necessary and sufficient criterion for two-mode Gaussian entanglement in terms of $\mathcal{D}^{\mathrm{sym}}$, placing on firm quantitative footing the intuition that the entanglement of a localized mode is encoded in the spatial support of its purification partner. We demonstrate the framework for wavepacket modes of a scalar quantum field in Minkowski spacetime (illustrating how the geometry of partner modes reveals the spatial structure of quantum correlations) and discuss extensions to multimode systems and mixed Gaussian states.
\end{abstract}

\maketitle

\section{Introduction}

In quantum field theory (QFT), the structure and distribution of entanglement in the vacuum and in more general states underlie a wide range of phenomena in foundational physics, from area-law behavior to the correlations associated with Hawking radiation \cite{witten, ReehSchlieder, SorkinArea1986,areaLaw1993, Hawking2005}. However, quantifying entanglement in QFT is particularly subtle. The infinite number of degrees of freedom, together with the absence of a preferred tensor-product decomposition of the Hilbert space into localized subsystems, poses significant challenges for characterizing entanglement between field modes. It is therefore useful to develop tools to identify and quantify correlations in such settings. In this work, we develop such tools for bosonic Gaussian systems, with particular emphasis on their application to quantum fields.

Gaussian states offer a tractable yet rich framework for exploring entanglement in continuous-variable quantum information and quantum field theory~\cite{GQMRev,Serafini2017,Adesso2014} since they can be characterized by their first and second statistical moments. Furthermore, they describe many physically relevant states, such as the ground and thermal states of quadratic Hamiltonians, as well as coherent and squeezed states. 

The study of entanglement and correlations for Gaussian states  has seen increasing development in recent years. In the context of QFT in a lattice, a recipe for finding the form of the most entangled modes respectively localized in two non-overlapping regions of space was provided in~\cite{Natalie1,NatalieUVIR,Natalie2,Natalie3}. However, the modes which contain the entanglement between the two regions happen to have a special form. This agrees with the results in \cite{ubiquitous}, where it was shown that randomly chosen pairs of modes of a quantum field localized in spacelike separated regions are typically not entangled. On the other hand, formulas for identifying purification partners for a Gaussian state in quantum field theory were first studied in~\cite{HottaPartner} and further developed in~\cite{Trevison2019}. This line of work has stimulated further investigation, including analyses of entanglement structures~\cite{tomitsuka_partner_2020,nambu_entanglement_2023,de_s_l_torres_entanglement_2023,osawa_final_2024,montes-armenteros_quantum_2025} and entanglement harvesting~\cite{trevison_pure_2018,hackl_minimal_2019,osawa_entanglement_2025}, as well as theoretical developments such as extensions to curved spacetimes \cite{Ribes-Metidieri:2025nfw,Ribes-Metidieri:2024vjn,Agullo:2024nxg}, fermionic modes~\cite{hackl_minimal_2019,hackl_bosonic_2021,Jonsson:2021lko}   and scenarios involving multiple modes~\cite{yamaguchi_quantum_2020,KojiCapacity,yamaguchi_quantum_2021,Agullo:2024har}.

More recently, we have developed a systematic framework centered on the use of complex phase spaces and restricted complex structures to study how bipartite correlations and entanglement are organized in both pure and mixed Gaussian states ~\cite{agullo_correlation_2025}. For pure Gaussian states, this framework provides a compact and elegant expression for the purification partner of any single-mode subsystem. The purification partner of a given mode $A$, which we denote by $A_p$, is a single mode distinct from and independent of $A$ that captures all correlations and entanglement associated with that mode. In other words, the reduced state of the two-mode system $(A, A_p)$ is pure and therefore uncorrelated with any other degrees of freedom.

Intuitively, the spatial support of the purification partner of a QFT mode contains information about where the correlations and entanglement of a mode are located in space. Namely, if the partner mode of $A$ is highly localized within a compact region of space, it is expected that the correlations of $A$ with the rest of the field are predominantly encoded in degrees of freedom associated with that region. This intuition has been used, for example, in discussions of the Hawking effect to argue that the correlations with a Hawking mode emitted by a black hole are localized inside the horizon~\cite{Wald:1975,Hotta2015,Agullo:2024nxg}. The main purpose of this article is to formalize this intuition and express it in quantitative terms.

Given  two modes $A$ and $B$ of a scalar field theory prepared in a pure Gaussian state, we introduce a quantity, denoted by $\mathcal{D}^{\mathrm{sym}}$, which quantifies the spatial overlap between $B$ and $A_p$, symmetrically combined with the overlap between $A$ and $B_p$. We prove $\mathcal{D}^{\mathrm{sym}}$ quantifies the correlations between $A$ and $B$, thereby linking their correlations to the profiles of their partner modes. This measure is invariant under local unitary operations within each subsystem. This measure, $\mathcal{D}^{\mathrm{sym}}$, extends naturally to subsystems consisting of an arbitrary finite number of modes.  Our analysis applies both to quantum mechanical systems and to quantum fields, although our primary focus is on the latter.

Regarding entanglement between two modes, we show that a necessary and sufficient condition for $A$ and $B$ to be entangled is that $\mathcal{D}^{\mathrm{sym}}$ exceeds a threshold value, which is itself determined by the purities of $A$, $B$ and of the combined system $(A,B)$. 

Finally, we show that for weakly entangled modes---which is typically the case in quantum field theory when $A$ and $B$ are local modes supported in non-overlapping spatial regions---$\mathcal{D}^{\mathrm{sym}}$ is directly related to the logarithmic negativity between $A$ and $B$. As an illustrative example, we present a numerical study of the entanglement between a family of local modes in Minkowski spacetime.
These results extend to mixed states.

We further show that  $\mathcal{D}^{\mathrm{sym}}$ is a symmetrized version of the correlation measure recently introduced in~\cite{osawa_entanglement_2025}. This symmetrization renders $\mathcal{D}^{\mathrm{sym}}$ invariant under the exchange of modes $A$ and $B$, a desirable property for a quantifier of correlations between them.

We formulate our results using a complex version of the classical phase space and the complex structures defined on it. These  tools have a clean geometric interpretation, and have the advantage of being manifestly coordinate independent, making it straightforward to verify invariance under local unitaries---which, in the phase-space language, reduces to invariance under local symplectic transformations.

Throughout this paper, we use natural units with $\hbar = c = 1$.

\section{Preliminaries}\label{sec:preliminaries}

With the aim of making this article self-consistent and of fixing the notation and terminology, this section reviews standard quantum-mechanical concepts for Gaussian states, with an emphasis on the use of the complexified phase space. We also review the partner formula, focusing in particular on the version obtained in~\cite{agullo_correlation_2025}

\subsection{Symplectic product in bosonic linear systems}
Consider a system with $N$ bosonic degrees of freedom, whose classical phase space $\Gamma$ is a $2N$-dimensional manifold space endowed with a symplectic two-form $\Omega_{ab}$. Its inverse $\Omega^{ab}$ satisfies $\Omega^{ac}\Omega_{cb}=\delta^a_b$, and defines the Poisson bracket $\{f,g\} \coloneqq \Omega^{ab} \, \partial_a f \, \partial_b g$ for functions $f$ and $g$ on $\Gamma$. For a linear system, global canonical coordinates, i.e., Darboux coordinates, $(r^1,\ldots,r^{2N})=(x^1,p^1,\ldots,x^N,p^N)$ exist, satisfying $\{r^i,r^j\}=\Omega^{ij}$, where the matrix with entries $\Omega^{ij}$ takes the block-diagonal form
\begin{align}
    \bigoplus_{N}
    \begin{pmatrix}
        0 & 1 \\
        -1 & 0
    \end{pmatrix}.
\end{align}
These global coordinates endow $\Gamma$ with the structure of a $2N$-dimensional real vector space; hence, elements of $\Gamma$ can be represented by a $2N$-component vector, which we denote by $\gamma^a$. Since $\Gamma$ is a linear space, it can be  identified with its own tangent space, which permits to introduce a ``symplectic product'' between vectors $\gamma^a, \gamma^{\prime a} \in \Gamma$ as 
\begin{align}
    \Omega(\gamma,\gamma') \coloneqq \Omega_{ab} \, \gamma^a \gamma^{\prime b}. 
\end{align}
The symplectic  structure $\Omega$ is an antisymmetric and non-degenerate bilinear form on $\Gamma$. Linear transformations preserving $\Omega$  correspond to linear canonical coordinate changes, called symplectic transformations, and form the symplectic group $Sp(2N,\mathbb{R})$.

To transition to the quantum theory, we introduce the vector of canonical operators $\hat{\bm{r}} \coloneqq (\hat{x}_1, \hat{p}_1, \ldots, \hat{x}_N, \hat{p}_N)$, which satisfies the canonical commutation relations
\begin{align}
    [\hat{r}^i,\hat{r}^j] = \ii \,\Omega^{ij}\,\hat{\openone}.
\end{align}
Such operators are referred to as ``Darboux operators''. Linear combinations of these operators, $c_i \hat{r}^i$ with $c_i \in \mathbb{C}$, form the set of elementary observables, from which more general observables can be constructed.  It is convenient to allow the coefficients $c_i$ to be complex, so that non-self-adjoint operators such as creation and annihilation operators are included in the discussion. This is the primary motivation for introducing the complexified phase space $\Gamma_{\mathbb{C}}$ below.

The symplectic structure $\Omega_{ab}$ and its inverse $\Omega^{ab}$ are extended to $\Gamma_{\mathbb{C}}$ by linearity, thereby defining the complexified symplectic product
\begin{align}
    \braket{\cdot,\cdot} \coloneqq \frac{1}{\ii}\,\Omega(\gamma^*,\gamma'),
\end{align}
where $*$ denotes complex conjugation (defined in the canonical coordinates that endow $\Gamma$ with its vector-space structure).  The complexified symplectic product $\braket{\cdot,\cdot}$ satisfies  all properties of a Hermitian inner product in $\Gamma_{\mathbb{C}}$, except positive definiteness.  

It is convenient to associate each vector $\gamma \in \Gamma_{\mathbb{C}}$  with a linear operator as follows
\begin{align}
    \gamma \in \Gamma_{\mathbb{C}} 
    \longleftrightarrow
    \hat{O}_\gamma \coloneqq \ii\braket{\gamma,\hat{\bm{r}}}
    = \Omega_{ij}\,\gamma^{* i}\,\hat{r}^j,\label{eq:correspondence_vector_operator}
\end{align} (sum over repeated indices is understood). In this expression, the contraction $\Omega_{ij}\,\gamma^{* i}$ plays the role of the coefficients $c_j$ mentioned above. 
With this definition, the commutator between  linear operators can be expressed in terms of a symplectic product as
\begin{align}
    [\hat{O}_\gamma,\hat{O}_{\gamma'}^\dag] = \braket{\gamma,\gamma'}\hat{\openone}.\label{Commutators_Symplectic_Product}
\end{align}

\subsection{Subsystems and symplectic projectors}
Adopting the algebraic approach in quantum theory, a subsystem can be characterized by a subalgebra of observables. According to the correspondence given in Eq.~\eqref{eq:correspondence_vector_operator} between phase space elements and operators, a subsystem corresponds to a symplectic subspace of the phase space. More precisely, any subspace $\Gamma_A \subset \Gamma_{\mathbb{C}}$ for which 
the restriction of $\Omega$ to $\Gamma_A$ constitutes a bona fide symplectic structure---namely, a non-degenerate two-form---defines a 
a subsystem $A$. In this case, $\Gamma_{\mathbb{C}}$ decomposes into a direct sum of symplectic subspaces, $\Gamma_{\mathbb{C}}=\Gamma_A\oplus \Gamma_{\bar{A}}$, where $\Gamma_{\bar{A}}$ denotes the symplectic orthogonal complement of $\Gamma_A$, defined as
\begin{align}
\Gamma_{\bar{A}}\coloneqq \{\gamma\in\Gamma_{\mathbb{C}}\mid \braket{\gamma,\gamma'}=0\quad \forall\gamma'\in\Gamma_A\}.
\end{align}

It is simple to prove that if $\Gamma_A$ is symplectic, so is  $\Gamma_{\bar{A}}$. So $\Gamma_{\bar{A}}$ itself defines another subsystem, referred to as the complement of $A$, denoted by $\bar{A}$. Accordingly, the symplectic structure $\Omega$ decomposes as $\Omega=\Omega_A\oplus \Omega_{\bar{A}}$, where $\Omega_A$ and $\Omega_{\bar{A}}$ denote the symplectic structures in $\Gamma_{A}$ and $\Gamma_{\bar{A}}$, respectively.

Due to the direct sum structure $\Gamma_{\mathbb{C}}=\Gamma_{A}\oplus \Gamma_{\bar{A}}$, any vector $\gamma\in\Gamma_{\mathbb{C}}$ can be uniquely decomposed as $\gamma=\xi+\xi'$ with $\xi\in\Gamma_A$ and $\xi'\in\Gamma_{\bar{A}}$. This allows the introduction of linear projectors $\Pi_{A}$ and $\Pi_{A}^\perp\coloneqq 1-\Pi_{A}$ onto the symplectic subspaces $\Gamma_A$ and $\Gamma_{\bar{A}}$, respectively. Since $\Pi_A\Pi_{A}^\perp=\Pi_A^\perp\Pi_A=0$, the symplectic product $\braket{\cdot,\cdot}$ satisfies a Pythagorean-type relation:
\begin{align}
\braket{\gamma,\gamma}=\braket{\Pi_A(\gamma),\Pi_A(\gamma)}+\braket{\Pi_A^\perp(\gamma),\Pi_A^\perp(\gamma)}\quad \forall\gamma\in\Gamma_{\mathbb{C}}.\label{eq:Pythagorean-type_decomposition}
\end{align}
Therefore, the quantity $\braket{\Pi_A(\gamma),\Pi_A(\gamma)}$ can be viewed as the contribution of subsystem $A$ to the total symplectic ``norm'' of $\gamma$ (recall that  $\braket{\cdot,\cdot}$ can either be positive, negative, or zero).

Although the symplectic projector $\Pi_A$ is defined independently of any particular choice of basis in $\Gamma_A$, it is often convenient to express it in terms of basis vectors. In what follows, we mainly focus on the case where $A$ corresponds to a single-mode subsystem. In this situation, one can always choose a basis $\{\gamma_A,\gamma_A^*\}$ of $\Gamma_A$ satisfying
\begin{align}
    \braket{\gamma_A,\gamma_A}=-\braket{\gamma_A^*,\gamma_A^*}=1,
\end{align}
or equivalently, $[\hat{a},\hat{a}^\dag]=\hat{\openone}$ with $\hat{a}\coloneqq \hat{O}_{\gamma_A}$. In this basis, the projector takes the simple form
\begin{align}
    \Pi_A(\cdot)=\gamma_A\braket{\gamma_A,\cdot}-\gamma_A^*\braket{\gamma_A^*,\cdot}.\label{eq:projector_in_terms_of_basis_vectors}
\end{align}
The generalization to a multi-mode subsystem can be found in Appendix \ref{app:multimode_and_mixed}.

\subsection{Gaussian states and the complex structure}
In the study of continuous-variable quantum systems, Gaussian states play a central role due to their analytical tractability and experimental relevance. This class encompasses many physically significant states, including squeezed and coherent states, as well as the ground and thermal states of quadratic Hamiltonians.

Gaussian states are completely characterized by the first and second moments of the canonical operators. When considering properties invariant under local unitary transformations---particularly correlations between subsystems---only the second moments are relevant, since the first moments can be eliminated by suitable local unitaries.

The complex structure offers an equivalent and often more convenient representation of the information contained in the covariance matrix, thereby providing a complete description of the correlation structure. The usefulness of both the phase-space approach and complex structures has been highlighted, for example, in \cite{hackl_minimal_2019,hackl_bosonic_2021,Jonsson:2021lko,agullo_correlation_2025}.
In what follows, we briefly review these notions and establish the notation used throughout this article.

To formalize these notions, we introduce a centered version of the linear operators introduced in the previous section, given by
\begin{align}\label{centeredOperator}
\hat{\overline{O}}_{\gamma} \coloneqq \hat{O}_{\gamma} - \mathrm{Tr}[\hat{\rho} \hat{O}_{\gamma}] \hat{\openone},
\end{align}
where $\hat{\rho}$ denotes the density operator describing a Gaussian state. This centering removes the contribution of the first moments and will serve as the basis for defining correlation quantities.

A state $\hat \rho$  defines a symmetric twice-covariant tensor $\sigma_{ab}$ on the complexified phase space $\Gamma_{\mathbb{C}}$, which captures the second-moment structure, as
\begin{align}
    \sigma(\gamma,\gamma') \coloneqq \mathrm{Tr}\left(\hat{\rho}\,\{{\hat{\overline{O}}_{\gamma}^{\dagger}, \hat{\overline{O}}_{\gamma'}^{\dagger}}\}\right), \quad \gamma,\gamma'\in\Gamma_{\mathbb{C}}.
    \label{eq:doubledaggerdef}
\end{align}
The adjoint conjugation of the linear operators appearing in this equation ensures that $\sigma$ is bilinear in the vectors $\gamma$ and $\gamma'$ (rather than anti-linear\footnote{With our conventions, the map $\gamma\mapsto\hat O_\gamma$ defined in \eqref{eq:correspondence_vector_operator} is anti-linear in $\gamma$, since $\langle\cdot,\cdot\rangle$ is Hermitian (anti-linear in its first argument). Taking adjoints therefore yields
$\hat O_\gamma^\dagger = -\,\hat O_{\gamma^{*}}$. As a result,
$
\{\hat O_\gamma^\dagger,\hat O_{\gamma'}^\dagger\}
=
\{\hat O_{\gamma^{*}},\hat O_{\gamma'^{*}}\}$. Because $\hat O_{\gamma^{*}}$ depends linearly on $\gamma$, the covariance tensor $\sigma(\gamma,\gamma')$ defined in \eqref{eq:doubledaggerdef} is genuinely complex-bilinear in its arguments, as stated.
}).

Having introduced the covariance tensor $\sigma$, we now define a corresponding linear map $J$ on $\Gamma_{\mathbb{C}}$ by raising one index of $\sigma_{ab}$ with the symplectic structure $\Omega$ :
\begin{align}
J^{a}_{\ b} \coloneqq -\, \Omega^{ac}\sigma_{cb}.\label{J_definition}
\end{align}

Because $\Omega$ is a fixed, invertible structure independent of the state, $J$ and $\sigma$ carry the same information. However, it is often more convenient to work with $J$, since it is a linear map, and admits a well-defined spectral structure---eigenvalues and eigenvectors (see \cite{agullo_correlation_2025} for further details and proofs omitted here). In contrast, $\sigma$, being a twice-covariant tensor, does not naturally possess such a spectral structure unless a specific basis is chosen.

The matrix $J$ has purely imaginary eigenvalues occurring in conjugate pairs $\pm \ii\nu_I$, with real numbers $\nu_I \ge 1$ for $I = 1, \ldots, N$ (see, e.g., \cite{serafini2017quantum,agullo_correlation_2025}). The real numbers $\nu_I$ are commonly referred to as the ``symplectic eigenvalues'' of $\sigma$. Importantly, a Gaussian state $\hat{\rho}$ is pure if and only if $\nu_I = 1$ for all $I$. Equivalently, this condition can be expressed as $J^2 = -\mathbb{I}$. A linear map that squares to minus the identity is called a complex structure. Therefore, a pure Gaussian state defines a complex structure in the classical phase space. In contrast, $J^2 < -\mathbb{I}$ holds for all mixed Gaussian state. Since any mixed Gaussian state can always be obtained as the reduced state of some pure Gaussian state, $J$ satisfying $J^2 < -\mathbb{I}$ can be regarded as the restriction of a complex structure to a subspace, and is therefore referred to as a restricted complex structure  (see e.g. \cite{hackl_bosonic_2021}).

\subsection{Symplectic invariants and entanglement for bipartite systems}\label{sec:twomodes}

The entanglement structure in many-body systems, including quantum fields, plays a crucial role in understanding their fundamental physical properties. Among the simplest yet most foundational quantities in this context is the entanglement between two independent modes, $A$ and $B$, in a Gaussian state. In this subsection, we briefly revisit several basic concepts, focusing on symplectic invariants that characterize a two-mode Gaussian state and the logarithmic negativity, a standard measure of bipartite entanglement.

Consider a bipartite system composed of modes $A$ and $B$. The union of the Darboux bases of the subsystems forms a Darboux basis of the total system. That is, for the Darboux bases $\hat{\bm{r}}_A = (\hat{x}_A, \hat{p}_A)$ and $\hat{\bm{r}}_B = (\hat{x}_B, \hat{p}_B)$ of the respective modes, the operator vector
\begin{align}
    \hat{\bm{r}} \coloneqq (\hat{x}_A, \hat{p}_A, \hat{x}_B, \hat{p}_B)
\end{align}
constitutes a Darboux basis of the composite system $AB$. The matrix elements of the restricted complex structure $J_{AB}$ of the state $\hat{\rho}_{AB}$ in this basis takes the form
\begin{align}\label{TwoModesNonHerm_CovMatrix}
J_{AB}=
\begin{pmatrix}
J_A & J_C\\
J_C^{\top} & J_B
\end{pmatrix}
\end{align}
where $J_A$, $J_B$, and $J_C$ are $2\times 2$ real matrices. 

There exist four quantities that are invariant under local symplectic transformations acting independently on modes $A$ and $B$, namely
\begin{align}\label{symplectic_invariants_J}
    \det J_{AB}, \quad \det J_A, \quad \det J_B, \quad \text{and } \det J_C .
\end{align}
This invariance follows from the fact that the determinant of a linear map is preserved under changes of coordinates. This statement holds not only for symplectic transformations, but for any change of basis in phase space\footnote{In contrast, because the covariance matrix is a rank-two tensor rather than a linear map, its determinant is not invariant under a general change of basis—although it is invariant under symplectic transformations, which have unit determinant. When using a Darboux basis, the determinants of $J$ and $\sigma$ coincide.}.
In this article, we will exploit the advantages of using a complex orthonormal basis in phase space (such a basis is naturally associated with annihilation and creation operators). It is therefore important that the determinants of $J$ remain invariant when changing from a real to a complex basis.

Entanglement between two modes, $A$ and $B$, is one of the most fundamental properties that remain invariant under local symplectic transformations. 
In general, for an arbitrary bipartite state, the positivity of the partially transposed density operator is a necessary condition for separability, known as the positive partial transpose (PPT) criterion \cite{Peres:1996dw,Horodecki:1996nc}. 
Conversely, non-positivity of the partial transpose is a sufficient condition for the presence of entanglement. 
Importantly, Simon~\cite{Simon2000} showed that the PPT criterion is both necessary and sufficient for the separability of two-mode Gaussian states. 

To explore this further, let us introduce the partially transposed (PT) restricted complex structure,
\begin{equation}
J^{T_A} =- \Omega \, T_A\, \sigma\, T_A\, ,
\end{equation}
where $T_A$ denotes the momentum flip on subsystem $A$, which in the canonical basis has the matrix representation
\begin{equation}
T_A = 
\begin{pmatrix}
    1 & 0 \\
    0 & -1
\end{pmatrix}.
\end{equation}
The PPT criterion then reduces to a simple inequality involving the smallest absolute value of the eigenvalues $\pm \ii \tilde{\nu}_{-}$ of $J^{T_A}$. Specifically, a two-mode Gaussian state is \textit{separable if and only if $\tilde{\nu}_{-} \geq 1$}.

A closely related entanglement measure is the \emph{logarithmic negativity} (LogNeg)~\cite{VidalNegativity, plenio05}, which we denote by $E_{\mathcal{N}}$. A nonzero value of the LogNeg indicates a violation of the PPT criterion. For a two-mode Gaussian state, the LogNeg can be computed directly from the partially transposed restricted complex structure and is given by
\begin{equation}\label{LogNeg}
      E_{\mathcal{N}} = \max\{0,-\log_{2} \tilde{\nu}_-\}.
\end{equation}

\subsection{Partner formula for pure Gaussian states}\label{sec:partner_review}

In this section, we recall the construction of the partner mode corresponding to a single-mode subsystem $A$ in a bosonic system described by a pure Gaussian state. The partner $A$, denoted by $A_p$, is the smallest subsystem independent of $A$ that encodes all correlations and entanglement with it. The partner-mode formula was first introduced in \cite{HottaPartner} and established for general pure Gaussian states in \cite{Trevison2019}. A reformulation based on  complex structures was later developed in \cite{hackl_minimal_2019, agullo_correlation_2025}. In the following, we adopt the formulation of \cite{agullo_correlation_2025}, which is manifestly basis-independent and symplectic invariant. The study of partner modes was further extended to mixed Gaussian states in \cite{agullo_correlation_2025}.

We begin by reviewing the notation and definitions introduced in \cite{agullo_correlation_2025}. Consider an $N$-mode bosonic system in a Gaussian state $\hat{\rho}$ and a subsystem $A$ composed of $N'(<N)$ modes. We say that $A$ is uncorrelated if and only if 
\begin{equation} \label{uncorr}
\mathrm{Tr}\left[\hat \rho\, \hat{\overline O}_{\gamma}\hat{\overline O}_{\gamma'}\right] = 0 
\end{equation}
for all $\gamma \in \Gamma_A$ and all $\gamma' \in \Gamma_{\bar A}$, where $\bar A$ denotes the symplectic orthogonal complement of $A$ in $\Gamma_{\mathbb{C}}$. The condition \eqref{uncorr} expresses the absence of mixed second moments between degrees of freedom in $\Gamma_A$ and its symplectic orthogonal complement $\Gamma_{\bar A}$, i.e., the vanishing of the cross-covariance tensor, $\sigma(\gamma,\gamma')=0$ for all $\gamma\in\Gamma_A$ and $\gamma'\in\Gamma_{\bar A}$.
If $A$ is uncorrelated, the state $\hat{\rho}$ factorizes into a product form as $\hat{\rho} = \hat{\rho}_A \otimes \hat{\rho}_{\bar A}$. We further say that a subsystem $A$ is correlated if it is not uncorrelated.

As shown in \cite{agullo_correlation_2025}, a subsystem $A$ is uncorrelated if and only if the complex structure $J$ leaves the associated symplectic subspace $\Gamma_A$ invariant, i.e., $J\Gamma_A = \Gamma_A$. When $\hat{\rho}$ is a pure Gaussian state and a single-mode subsystem $A$ is correlated, it was shown in \cite{agullo_correlation_2025} that the subspace associated with the partner $A_p$ is given by
\begin{equation}\label{singl_corr_partner_formula}
\Gamma_{A_p} = \Pi^\perp_A (J \Gamma_A)\, .
\end{equation} 
This follows from the facts that $\Gamma_A \oplus \Gamma_{A_p} = \Gamma_A + J\Gamma_A$ and that $J^2 = -\mathbb{I}$ holds for a pure Gaussian state $\hat{\rho}$. With this, it is straightforward to show that $J (\Gamma_A \oplus \Gamma_{A_p})=\Gamma_A \oplus \Gamma_{A_p}$, so the system $\Gamma_A \oplus \Gamma_{A_p}$ is uncorrelated. Therefore, the partner subsystem $A_p$ contains all correlations with $A$.

Despite the concise and basis-independent expression in  Eq.~\eqref{singl_corr_partner_formula}, it is often convenient to choose a basis for the partner-mode subsystem. Let $\gamma_A$ be any vector in $\Gamma_A$ satisfying $\braket{\gamma_A, \gamma_A} = 1$. Then, the vector
\begin{equation}\label{PartnerPosBasisVector}
\gamma_{A_p} = \frac{1}{\sqrt{\det J_A - 1}}\, \Pi_A^{\perp}(J \gamma_A^{*})
\end{equation}
together with its complex conjugate form an orthonormal basis in $\Gamma_{A_p}$---orthonormal in the following sense:  $\langle \gamma_{A_p}, \gamma_{A_p} \rangle = 1$, $\langle \gamma^*_{A_p}, \gamma^*_{A_p} \rangle = -1$ and $\langle \gamma_{A_p}, \gamma^*_{A_p} \rangle = 0$ (the prefactor $(\sqrt{\det J_A - 1})^{-1}$ ensures  normalization).

As an illustration, we provide a simple example for the calculation of the partner mode:

\begin{widetext}

\begin{tcolorbox}[colframe=gray, colback=yellow!2, breakable]\label{ExampleHO1}
\textbf{Example: 2-mode squeezed vacuum combined with a third mode through a ``beam splitter''}

Consider three harmonic oscillator $ABC$, all with the same  mass \( m \) and frequency \( w_0 \). 

The following four vectors 
\begin{align}
\gamma_1 &= \sqrt{m w_0}\, \begin{pmatrix} 0\\ 1\\ 0\\ 0 \\0\\0\end{pmatrix}, \
\gamma_2 = - \sqrt{\frac{1}{m w_0}}\, \begin{pmatrix} 1\\ 0\\ 0\\ 0 \\0\\0\end{pmatrix}, \
\gamma_3 = \sqrt{m w_0}\, \begin{pmatrix} 0\\ 0\\ 0\\ 1 \\0\\0\end{pmatrix}, \ 
\gamma_4 = - \sqrt{\frac{1}{m w_0}}\, \begin{pmatrix} 0\\ 0\\ 1\\ 0 \\0\\0\end{pmatrix}\nonumber\\
\gamma_5 &= \sqrt{m w_0}\, \begin{pmatrix} 0\\ 0\\ 0\\ 0 \\0\\1\end{pmatrix}, \ 
\gamma_6 = - \sqrt{\frac{1}{m w_0}}\, \begin{pmatrix} 0\\ 0\\ 0\\ 0 \\1\\0\end{pmatrix}\nonumber 
\end{align}
form a Darboux basis in the classical phase space. The  operators associated with these basis vectors via the correspondence (\ref{eq:correspondence_vector_operator}) are the (dimensionless) self-adjoint operators 
\[
\sqrt{m w_0}\, \hat x_A, \quad \sqrt{\frac{1}{m w_0}}\, \hat p_A, \quad \sqrt{m w_0}\, \hat x_B, \quad \sqrt{\frac{1}{m w_0}}\, \hat p_B, \quad
\sqrt{m w_0}\, \hat x_C, \quad \sqrt{\frac{1}{m w_0}}\, \hat p_C,
\]
where $\hat x_I$ and $\hat p_I$, $I=A,B,C$ represent the standard position and momentum operators. In this basis, the covariance matrix of the vacuum state---the ground state of the Hamiltonian---is $\sigma_0= \mathbb{I}_{6\times6}$. This covariance matrix describes a pure state---the associated complex structure $(J_0)^{a}_{\ b}= -\Omega^{ad}\sigma_{db}$, has eigenvalues $\pm \, \ii$. The eigenvectors of $J_0$ are
$$ \gamma_A= \frac{1}{\sqrt{2}}(\gamma_1 -\ii \gamma_2) \quad;\quad  \gamma_B= \frac{1}{\sqrt{2}}(\gamma_3 -\ii \gamma_4) \quad;\quad  \gamma_C= \frac{1}{\sqrt{2}}(\gamma_5 -\ii \gamma_6) \, , $$
together with their complex conjugated vectors (the operators associated with these complex vectors are, respectively, the annihilation and creation operators 
$\hat{\bm A}= (\hat{a}_A, \hat{a}^{\dagger}_A, \hat{a}_B, \hat{a}^{\dagger}_B,\hat{a}_C, \hat{a}^{\dagger}_C)$).

Subsystem $A$ is characterized by the symplectic subspace $\Gamma_{A}$ of  $\Gamma_{\mathbb{C}}$, given by
\[
\Gamma_{A} = \mathrm{span}\left[ \gamma_A,\gamma_A^*\right].
\]
That $\Gamma_A$ is spanned by an eigenvector of $J_0$ and its complex conjugate automatically implies that $A$ is an uncorrelated subsystem---since $J_0\Gamma_A$  is obviously equal to $\Gamma_A$.

The symplectic projector $\Pi_A$ onto $\Gamma_{A}$ can be written as
\[
\Pi_A (\cdot)= \gamma_A \langle \gamma_A, \cdot \rangle - \gamma_A^* \langle \gamma_A^*, \cdot \rangle.
\]

Now, we generate a new Gaussian state by applying a two-mode squeezing transformation between oscillators $A$ and $B$ (with squeezing intensity $r$ and squeezing angle $\phi=0$) to the vacuum state followed by a beam splitter transformation between modes $B$ and $C$. Mathematically, these transformations can be implemented via a symplectic transformation in the classical phase space given, respectively, by 

\begin{align}
S_{AB}=
\begin{pmatrix}
 \cosh r & 0 & \sinh r & 0 & 0 & 0 \\
 0 & \cosh r & 0 & -\sinh r & 0 & 0 \\
 \sinh r & 0 & \cosh r & 0 & 0 & 0 \\
 0 & -\sinh r & 0 & \cosh r & 0 & 0 \\
 0 & 0 & 0 & 0 & 1 & 0 \\
 0 & 0 & 0 & 0 & 0 & 1 \\
\end{pmatrix}\, ,
\end{align}
and 
\begin{align}
M_{BC}=
\begin{pmatrix}
1 & 0 & 0 & 0 & 0 & 0 \\
 0 & 1 & 0 & 0 & 0 & 0 \\
 0 & 0 & \cos \theta & 0 & \sin \theta & 0 &  \\
  0 & 0 &0 & \cos \theta & 0 & \sin \theta  \\
 0 & 0 & -\sin \theta & 0 & \cos \theta & 0 \\
 0 & 0 & 0 & -\sin \theta & 0 & \cos \theta  \\
\end{pmatrix} \, .
\end{align}
The covariance matrix of the resulting state (written in the Darboux basis specified above) is
\[
\sigma \;=\; M_{BC}\, S_{AB}\, \sigma_0 \, S_{AB}^{T} \, M_{BC}^{T} \, .
\]
This covariance matrix describes a pure state---the associated complex structure
\[
J^{a}{}_{b} = -\, \Omega^{ad}\sigma_{db}
\]
has eigenvalues $\pm\, i$. The eigenvectors of $J$ are the complex vectors 

\begin{align}
    e_{1} =&  \frac{\csc \theta  \coth r}{\sqrt{2}} \, \gamma_A + -\frac{\cos (2 \theta ) \csc \theta \tanh
   r}{\sqrt{2}} \gamma_A^* +\frac{\cot \theta }{\sqrt{2}}(\gamma_B -\gamma_B^{*}) + \frac{1}{\sqrt{2}}(\gamma_C +\gamma_C^{*}) \, ,\\
    e_{2} =&  \frac{\ii \csc \theta  \coth r}{\sqrt{2}} \, \gamma_A 
   -\frac{\ii \csc \theta  \tanh r}{\sqrt{2}} \,\gamma_A^{*} + \frac{\ii \cot \theta
   }{\sqrt{2}}(\gamma_B -\gamma_B^{*}) - \frac{\ii}{\sqrt{2}}(\gamma_C -\gamma_C^{*}) \, ,\\
    e_{3} =& -\sqrt{2} \cos \theta  \tanh r\,  \gamma_A^{*} + \sqrt{2}\, \gamma_B \, .
\end{align}
together with their complex conjugated. The subspace $\Gamma_A$ is not  left invariant by the action of $J$, meaning that the subsystem $A$ is correlated (and entangled) when the system is prepared in the Gaussian state defined by $J$. We are interested in finding the purification partner of $A$. We expect this partner to be a combination of the modes $B$ and $C$ because, on the one hand, the two-mode squeezing transformation entangles modes $A$ and $B$, making $B$ the partner of $A$, but on the other hand the beam-splitter transformation mixes modes $B$ and $C$.

We compute the partner mode of $A$ by applying the linear map $\Pi_A^{\perp} J$ to the basis vectors in $\Gamma_A$ and normalizing. This action can be evaluated directly from the eigenvectors of $J$ together with those of $\Pi_A$. The result is
\[
\Gamma_{A_p} \;=\;
\mathrm{span}\!\left[
\, i\!\left(\cos\theta\, \gamma_B - \sin\theta\, \gamma_C\right),
\; -\, i\!\left(\cos\theta\, \gamma_B^{*} - \sin\theta\, \gamma_C^{*}\right)
\right].
\]

As expected, the partner of $A$ is a combination of modes $B$ and $C$ with weights $\cos\theta$ and $\sin\theta$. As a consistency check, when $\theta = 0$---for which the beam-splitter transformation is the identity---the partner of $A$ is the mode $B$, while for $\theta = (n + \tfrac{1}{2})\pi$ with $n \in \mathbb{N}$---corresponding to a completely reflective beam-splitter transformation---the partner of $A$ is the mode $C$.
 
\end{tcolorbox}
\end{widetext}

\section{Overlap formula}\label{sec:Overlap}
In this section, we begin by introducing a notion of overlap between two given modes. We then use the overlap of a mode $A$ with the partner of a mode $B$---and vice versa---to define a quantifier of the correlations between $A$ and $B$.

\subsection{Overlap between modes}\label{sec:overlap_definition}
Consider two modes $X$ and $Y$, which could be  subsystems of a larger system. The formalism described so far permits us to characterize, in a geometric manner, when the subsystems $X$ and $Y$ describe physically independent degrees of freedom: $X$ and $Y$ are independent if and only if the associated symplectic subspaces $\Gamma_X$ and $\Gamma_Y$ are symplectically orthogonal, namely
\[
\langle \gamma_X , \gamma_Y \rangle = 0, \qquad 
\forall\, \gamma_X \in \Gamma_X,\ \gamma_Y \in \Gamma_Y.
\]
This is so because symplectic orthogonality guarantees that all observables in subsystem $X$ commute with all observables in subsystem $Y$, since commutators are determined by symplectic products [see Eq.~\eqref{Commutators_Symplectic_Product}], thereby proving their independence. (For systems with finitely many degrees of freedom, the commutativity of the subalgebras of observables associated with each subsystem is equivalent to the familiar tensor-product factorization of the Hilbert space.)

This motivates us to define a geometric measure of ``non-orthogonality'' between subsystems.  
Let $\Pi_X$ denote the symplectic projector onto $\Gamma_X$.  
Let  $\gamma_Y$ be any vector in $\Gamma_Y$ satisfying the normalization condition $\langle \gamma_Y, \gamma_Y \rangle = 1$. We define
\begin{align}
    \mathcal{D}_{XY}
    \coloneqq 
    \langle \Pi_X(\gamma_Y),\, \Pi_X(\gamma_Y) \rangle,
    \label{eq:definition_overlap}
\end{align}
which corresponds to the symplectic norm of the projection of $\gamma_Y$ onto $\Gamma_X$.

Adopting a symplectic orthonormal basis $\{\gamma_X, \gamma_X^{*}\}$ of $\Gamma_X$ and writing the projector $\Pi_X$ in this basis, $\mathcal{D}_{XY}$ can be expressed as
\begin{align}
    \mathcal{D}_{XY}
    = 
    |\langle \gamma_X , \gamma_Y \rangle|^2
    - |\langle \gamma_X^{*} , \gamma_Y \rangle|^2.
    \label{eqOverlap}
\end{align}
Using the standard identities for the symplectic product,
\[
\langle \gamma , \gamma' \rangle^{*}
= - \langle \gamma^{*} , \gamma'^{*} \rangle
= \langle \gamma' , \gamma \rangle,
\]
we find that the overlap is symmetric:
\begin{align}
    \mathcal{D}_{XY}
    = 
    \mathcal{D}_{YX}.
    \label{eq:symmetric_property_of_D}
\end{align}

Importantly, $\mathcal{D}_{XY}$ is independent of the choice of basis in $\Gamma_X$, since the projector $\Pi_X$ is basis independent. By the symmetry property \eqref{eq:symmetric_property_of_D}, the same holds for $\Gamma_Y$. Since any two unit vectors in $\Gamma_Y$ are related by a local symplectic transformation on $Y$, invariance under such transformations ensures that $\mathcal{D}_{XY}$ is independent of the particular unit-norm vector $\gamma_Y \in \Gamma_Y$ chosen initially. Since $\mathcal{D}_{XY}$ is invariant under independent local symplectic transformations on $X$ and $Y$, it can be interpreted as a notion of \emph{overlap} between the subsystems $X$ and $Y$ themselves. It is immediate to see that if subsystems $X$ and $Y$ are independent, then $\mathcal{D}_{XY}=0$ since all vectors in $\Gamma_Y$ have vanishing projection onto $\Gamma_X$, and vice versa.  
The converse, however, is not true. This is because $\mathcal{D}_{XY}$ is not positive definite, and it can vanish in situations in which  the two terms in Eq.~\eqref{eqOverlap} are different from zero. 

\begin{tcolorbox}[colframe=gray, colback=yellow!2, breakable]\label{ExampleHO1}
\textbf{Example.} 
In the phase space of two harmonic oscillators, consider the following vectors:
\begin{align}
\gamma_1 = \begin{pmatrix}
0\\
\sqrt{m \omega_0}\\
0\\
0
\end{pmatrix},
\ 
\gamma_2 &= 
-\,
\begin{pmatrix}
\sqrt{\frac{1}{m \omega_0}}\\
0\\
0\\
0
\end{pmatrix}, \nonumber \\
\gamma_3 = 
\begin{pmatrix}
0\\
\sqrt{m \omega_0 }\\
0\\
\sqrt{m \omega_0 }
\end{pmatrix},
\
\gamma_4 &= 
-\,
\begin{pmatrix}
0\\
\sqrt{m \omega_0 }\\
\sqrt{\frac{1}{m \omega_0}}\\
0
\end{pmatrix}. \nonumber
\end{align}

Define the subsystems
\[
\Gamma_X = \mathrm{span}\{\gamma_1,\gamma_2\}, 
\qquad
\Gamma_Y = \mathrm{span}\{\gamma_3,\gamma_4\}.
\]

These subsystems satisfy $\mathcal{D}_{XY} = 0$, but they are \emph{not} independent.  
This can be seen by noting that the vector
\[
\frac{1}{\sqrt{2}} \left( \gamma_3 - \ii\, \gamma_4 \right) \in \Gamma_Y
\]
has unit symplectic norm, while its projection onto $\Gamma_X$ has vanishing symplectic norm.  
From Eq.~\eqref{eqOverlap} it then follows that $\mathcal{D}_{XY} = 0$.

That $X$ and $Y$ are nevertheless not independent can be verified directly by finding a vector in $\Gamma_X$ and a vector in $\Gamma_Y$ that are not symplectically orthogonal.  
For instance,
\[
\langle \gamma_2 , \gamma_3 \rangle \neq 0.
\]
\end{tcolorbox}

Although our main focus in this paper is on  single-mode systems $X$ and $Y$, the definition of  $\mathcal{D}_{XY}$ extends to subsystems made of any finite number of modes (see Appendix~\ref{app:multimode_and_mixed} for further details).

\subsection{Correlation and overlap}
The overlap defined in the previous subsection is purely kinematic, in the sense that it does not depend on the quantum state in which the system is prepared. Consequently, there is no direct relation between the overlap of two modes and their correlations, since the latter are determined by the state.

In this subsection, we argue that the overlap between a mode $A$ \emph{and the partner of} $B$ can be used to quantify the correlations between $A$ and $B$. Unlike the  overlap between $A$ and $B$, this quantity \emph{does} depend on the state, because the very notion of a partner mode is state-dependent.

Let $A$ and $B$ be two \emph{independent} single-mode subsystems within a larger system prepared in a pure Gaussian state $\hat \rho$. As reviewed in Section~\ref{sec:partner_review}, each mode admits a unique single-mode purification partner, denoted $A_p$ and $B_p$, respectively. Although $A$ and $B$ are independent---and therefore have vanishing overlap---the overlaps of $A_p$ with $B$, or of $B_p$ with $A$, may be nonzero.

\begin{tcolorbox}[colframe=black!35, colback=blue!2]
\begin{lemma} 

Let $A$ and $B$ be two independent single-mode subsystems of a system prepared in a pure Gaussian state. If $A$ and $B$ are uncorrelated with each other, then
\begin{equation}
    \mathcal{D}_{A_p B} = 0 = \mathcal{D}_{A B_p}.
\end{equation}

\end{lemma}
\end{tcolorbox}

\begin{proof}

The o $A$ and $B$ are correlated if and only if 
\[
\sigma(\gamma_A,\gamma_B)\neq 0
\]
for \emph{some} $\gamma_A\in\Gamma_A$ and $\gamma_B\in\Gamma_B$, where $\sigma$ is the covariance matrix of the pure Gaussian state $\hat \rho$. This quantity can be written as
\begin{equation}
    \sigma(\gamma_A,\gamma_B)
    = -\ii \langle \gamma_A^{*}, J\gamma_B \rangle.
    \label{eq:sigmaJ}
\end{equation}
Thus, the absence of correlations implies
\begin{equation}\label{unc}
     \langle \gamma_A^{*}, J\gamma_B \rangle
    = 0
\end{equation}
for all $\gamma_A \in \Gamma_A$ and $\gamma_B \in \Gamma_B$.

We decompose $J\gamma_B$ into its components within and out of $\Gamma_B$:
\begin{equation}
    J\gamma_B = \Pi_B ( J\gamma_B ) + \Pi_B^{\perp} ( J\gamma_B ).
\end{equation}
Independence of $A$ and $B$ implies
\[
\langle \gamma_A^{*},\, \Pi_B(J\gamma_B) \rangle = 0, 
\] since $\Gamma_A$ is symplectically orthogonal to $\Gamma_B$. 
Using this, condition~\eqref{unc} reduces to
\begin{equation}
    \langle \gamma_A^{*},\, \Pi_B^{\perp}( J\gamma_B ) \rangle = 0,
\end{equation}
for all $\gamma_A\in\Gamma_A$ and $\gamma_B\in\Gamma_B$. But $\Pi_B^{\perp} J\gamma_B$ and its conjugate span the partner mode of $B$, i.e.~$\Gamma_{B_p}$. Hence the above is equivalent to saying that $\Gamma_A$ is symplectically orthogonal to $\Gamma_{B_p}$, from which
\[
\mathcal{D}_{A B_p} = 0
\]
automatically follows.

The proof that $\mathcal{D}_{A_p B}=0$ is identical after exchanging the roles of $A$ and $B$. 

\end{proof}
\medskip

The contrapositive of this lemma implies that nonzero values of 
$\mathcal{D}_{A B_p}$ or $\mathcal{D}_{A_p B}$ signal the presence of correlations between $A$ and $B$.\footnote{The converse is not necessarily true: $\mathcal{D}_{A B_p}=0$ and $\mathcal{D}_{A_p B}=0$ do not in general imply the absence of correlations. This is the case, for example, for a two-mode system with non-zero position–position correlations while the momentum–momentum and position-momentum correlations vanish.
}

It is important to emphasize that $\mathcal{D}_{A B_p}$ coincides with a quantity recently  proposed in Ref.~\cite{osawa2025entanglement}, and used there as a necessary condition for entanglement. 

Our article adds to this discussion by providing a geometric and manifestly basis-independent interpretation of $\mathcal{D}_{A B_p}$. More importantly, in the remainder of this section we build on this quantity to establish a necessary and sufficient condition for entanglement, based on a suitable extension of the overlap $\mathcal{D}_{A B_p}$.

An important observation is that $\mathcal{D}_{A_p B}$ is not symmetric under the interchange of $A$ and $B$, i.e.\ $\mathcal{D}_{A_p B} \neq \mathcal{D}_{B_p A}$. This asymmetry reduces the appeal of $\mathcal{D}_{A_p B}$ as a measure of correlations. The asymmetry, however, can be readily eliminated by introducing a symmetrized version of the overlap, which involves the four relevant modes $A$, $B$, $A_p$, and $B_p$, and is defined as
\begin{align}\label{eqSymOverlap}
 \mathcal{D}^{\text{sym}} 
    &\coloneqq \mathcal{D}_{A_p B} + \mathcal{D}_{B_p A} \\
    &= \langle \Pi_{A_p}(\gamma_B), \Pi_{A_p}(\gamma_B) \rangle 
     + \langle \Pi_{B_p}(\gamma_A), \Pi_{B_p}(\gamma_A) \rangle \, , \nonumber
\end{align}
where in this expression $\gamma_B$ and $\gamma_A$ are any unit-symplectic norm vectors in $\Gamma_B$ and $\Gamma_A$, respectively.  We will refer to  $\mathcal{D}^{\text{sym}}$  as the {\em symmetric overlap}.

Using the partner-mode formula in Eq.~\eqref{singl_corr_partner_formula}, together with the local symplectic invariance of the overlap expression in Eq.~\eqref{eq:symmetric_property_of_D}, it follows directly that $\mathcal{D}^{\text{sym}}$ is itself invariant under local symplectic transformations acting on $\Gamma_A$ and $\Gamma_B$. The following proposition provides an explicit expression for $\mathcal{D}^{\text{sym}}$ in terms of the symplectic invariants.

\begin{tcolorbox}[colframe=black!35,
colback=blue!2]

\begin{proposition}
The symmetric overlap $\mathcal{D^{\text{sym}}}$ is related to the symplectic invariants $\det J_C$, $\det J_A$, and $\det J_B$ in the following way,
\begin{equation}\label{OverlapFormulaDeterminants}
\mathcal{D^{\text{sym}}}:=\left(\frac{1}{\det J_A-1}+\frac{1}{\det J_B-1}\right)(-\det J_C).
\end{equation}
where $J_A$ and $J_B$ are the $2\times 2$ restricted complex structure matrices of the reduced state for each subsystem $A$ and $B$. $J_C$ is a sub-matrix encoding the correlations present between the two mode system.
\end{proposition}

\end{tcolorbox}

\begin{proof}
Consider a pure Gaussian state $\hat\rho$ and let $J$ be its associated complex structure. It is convenient to work with an orthonormal basis in $\Gamma_A\oplus\Gamma_B$,
$\{\gamma_A,\gamma_A^{*},\gamma_B,\gamma_B^{*}\}$, satisfying
\begin{eqnarray}
\langle\gamma_A,\gamma_A\rangle &=& 1
= \langle\gamma_B,\gamma_B\rangle \,,\nonumber\\
\langle\gamma_A^{*},\gamma_A^{*}\rangle &=& -1
= \langle\gamma_B^{*},\gamma_B^{*}\rangle \,,
\end{eqnarray}
with all other products vanishing.

In this basis, the components of $J_{AB}$ can be written as
\begin{widetext}
\begin{equation}\label{TwoModesNonHerm_CovMatrix}
J_{AB}=
\begin{pmatrix}
J_A & J_C\\
J_C^{\top} & J_B
\end{pmatrix}
=
\begin{pmatrix}
\langle \gamma_A, J\gamma_A\rangle
&
\langle \gamma_A, J\gamma_A^{*}\rangle
&
\langle \gamma_A, J\gamma_B\rangle
&
\langle \gamma_A, J\gamma_B^{*}\rangle
\\
-\langle \gamma_A^{*}, J\gamma_A\rangle
&
-\langle \gamma_A^{*}, J\gamma_A^{*}\rangle
&
-\langle \gamma_A^{*}, J\gamma_B\rangle
&
-\langle \gamma_A^{*}, J\gamma_B^{*}\rangle
\\
\langle \gamma_B, J\gamma_A\rangle
&
\langle \gamma_B, J\gamma_A^{*}\rangle
&
\langle \gamma_B, J\gamma_B\rangle
&
\langle \gamma_B, J\gamma_B^{*}\rangle
\\
-\langle \gamma_B^{*}, J\gamma_A\rangle
&
-\langle \gamma_B^{*}, J\gamma_A^{*}\rangle
&
-\langle \gamma_B^{*}, J\gamma_B\rangle
&
-\langle \gamma_B^{*}, J\gamma_B^{*}\rangle
\end{pmatrix}.
\end{equation}
\end{widetext}

Using the properties of the symplectic product,
\(
\langle \gamma,\gamma'\rangle^{*}
= -\langle \gamma^{*},\gamma'^{*}\rangle
= \langle \gamma',\gamma\rangle
\),
and the fact that $J$ is real (so that $J\gamma_I^{*}=(J\gamma_I)^{*}$), one finds
\begin{align}\label{JSympDet2modes}
\det J_I &=
\bigl|\langle\gamma_I, J\gamma_I\rangle\bigr|^{2}
-
\bigl|\langle\gamma_I, J\gamma_I^{*}\rangle\bigr|^{2},
\qquad I=A,B,\\
\det J_C &=
\bigl|\langle\gamma_A, J\gamma_B\rangle\bigr|^{2}
-
\bigl|\langle\gamma_A, J\gamma_B^{*}\rangle\bigr|^{2}.
\end{align}

The following basis vector, which, together with its complex conjugate, span the partner-mode subsystems:
\begin{equation}\label{PartnerPosBasisVector}
\gamma_{I_p}=\frac{1}{\sqrt{\det J_I-1}}\Pi_I^{\perp}\bigl(J\gamma_I^{*}\bigr),
\qquad I=A,B\, .
\end{equation}

We now express the overlap in terms of basis vectors for modes $A$ and $B$ and their partners:
\begin{align}\label{eqOverlapPartner2}
\mathcal{D}_{B_pA}
=&
\langle \Pi_{B_p}(\gamma_A), \Pi_{B_p}(\gamma_A) \rangle \nonumber\\
=&
\bigl|\langle \gamma_{B_p},\gamma_A \rangle\bigr|^{2}
-
\bigl|\langle \gamma_{B_p}^{*},\gamma_A \rangle\bigr|^{2}.
\end{align}

Using the  vector $\gamma_{B_p}$ from
Eq.~\eqref{PartnerPosBasisVector} and its complex conjugate, we obtain (up to normalization)
\begin{align}
\langle \gamma_{B_p},\gamma_A \rangle
&\propto
\langle \Pi_B^{\perp}(J\gamma_B^{*}),\gamma_A\rangle
=
\langle J\gamma_B^{*},\Pi_B^{\perp}(\gamma_A)\rangle
=
\langle \gamma_A, J\gamma_B^{*}\rangle^{*}, \nonumber\\
\langle \gamma_{B_p}^{*},\gamma_A \rangle
&\propto
\langle \Pi_B^{\perp}(J\gamma_B),\gamma_A\rangle
=
\langle J\gamma_B,\Pi_B^{\perp}(\gamma_A)\rangle
=
\langle \gamma_A, J\gamma_B\rangle^{*},\nonumber
\end{align}
where we have used $\Pi_B^{\perp}(\gamma_A)=\gamma_A$ due to the orthogonality of $\Gamma_A$ and $\Gamma_B$.

Comparing these expressions with the symplectic invariant determinants in
Eq.~\eqref{JSympDet2modes} and restoring the normalization factor, we obtain
\begin{equation}\label{OverlapBpA_DetC}
\mathcal{D}_{B_pA}=\frac{-\det J_C}{\det J_B-1}.
\end{equation}

Consequently, the symmetrized overlap reads
\begin{equation}\label{OverlapFormulaDeterminants2}
\mathcal{D}^{\mathrm{sym}}
=
\left(
\frac{1}{\det J_B-1}
+
\frac{1}{\det J_A-1}
\right)(-\det J_C) \, ,
\end{equation}
which completes the proof.
\end{proof}
\subsection{Extension to mixed states}

So far we have assumed that $A$ and $B$ are subsystems of a larger system prepared in a pure Gaussian state. This assumption allowed us to define the purification partners of $A$ and $B$, from which the overlaps and ultimately $\mathcal{D}^{\mathrm{sym}}$ were constructed. 

However, inspection of Eq.~\eqref{OverlapFormulaDeterminants2} shows that $\mathcal{D}^{\mathrm{sym}}$ depends only on the reduced state of the pair $(A,B)$, and not on the manner in which this subsystem is purified within a larger system. In other words, $\mathcal{D}^{\mathrm{sym}}$ is insensitive to the details of any particular purification of $\hat{\rho}_{AB}$.

This observation makes it possible to extend the use of $\mathcal{D}^{\mathrm{sym}}$, as defined in Eq.~\eqref{OverlapFormulaDeterminants2}, to mixed states. The only requirement is that the reduced state $\hat{\rho}_{AB}$ be Gaussian. Furthermore, $\mathcal{D}^{\mathrm{sym}}$ can still be interpreted as quantifying the overlap between the purification partners of $A$ and $B$: one may, if desired, construct an explicit purification of $\hat{\rho}_{AB}$ by embedding it into a larger system. Yet, because $\mathcal{D}^{\mathrm{sym}}$ is independent of the purification chosen, such a construction is not necessary.

In simpler terms, the expression \eqref{OverlapFormulaDeterminants2} for $\mathcal{D}^{\mathrm{sym}}$ applies directly to mixed Gaussian states $\hat{\rho}_{AB}$ and may be viewed as a symmetric measure of the overlap between the modes and the purification partners that arise in any purification of the system $(A,B)$.

\subsection{Necessary and sufficient condition for entanglement between local modes\label{subsec:NecessarySufficientOverlap}}


For a two-mode subsystem $AB$ prepared in a 
Gaussian state $\hat \rho_{AB}$,  PPT is a necessary and sufficient condition for the separability, as summarized in Section \ref{sec:twomodes}. This PPT criterion reduces to a simple inequality for the smallest symplectic eigenvalue $\tilde \nu_{-}$ of the PT  complex structure of the reduced subsystem containing $AB$. Thus, we have that {\em a two-mode Gaussian state is separable (i.e. not entangled) if and only if $$\tilde \nu_{-} \geq 1 \, . $$}

The goal of this section is to recast this criterion in terms of an inequality for the symmetric overlap  $\mathcal{D}^{sym}$.

We begin from the following necessary condition for separability: 

\begin{tcolorbox}[colframe=gray, colback=white, breakable]
\begin{lemma}[Gaussian separability]
Two-mode Gaussian states with $\mathcal{D}^{sym} \leq 0$ are separable. Conversely, $\mathcal{D}^{sym} >0$ is a necessary condition for the two modes to be entangled.
\end{lemma}
\end{tcolorbox}
This follows directly from applying  \eqref{OverlapFormulaDeterminants} to Simon's separability lemma\footnote{{\bf Simon's separability lemma~\cite{Simon2000}:} Two-mode Gaussian states, whose correlations satisfy $\det J_C \geq 0$, are separable.} and noticing that $\det J_A \geq 1$ and $\det J_B \geq 1$. These determinants only equal one when the reduced state for each subsystem is pure, in which case there are no correlations at all.

We now evaluate the smallest symplectic eigenvalue, $\tilde{\nu}_{-}$, of the partially transposed restricted complex structure $J^{T_A}_{AB}$ for the two-mode subsystem. Following the arguments in \cite{serafini2017quantum}, we investigate the conditions on $\mathcal{D}^{sym}$ under which $\tilde{\nu}_{-} < 1$ holds.

First, recall that for a two-mode subsystem, the symplectic eigenvalues of the quantum state $\nu_{\pm}$ can be expressed in terms of the symplectic invariants~\cite{serafini2017quantum} as:
\begin{equation}
\nu_{\pm}^{2} =\frac{\Delta \pm \sqrt{\Delta^2 -4 \det J_{AB}}}{2}
\end{equation}
where 
\begin{align}
    \Delta :=& \det J_{A}+\det J_B+2\det J_C = \nu_+^2+\nu_-^2\quad \text{and }\nonumber\\
    \det J_{AB}=&\nu_+^2\nu_-^2 \, .
\end{align}
The positivity of the density  operator implies  $\nu_{\pm} \geq 1$. This condition is equivalent to the following set of inequalities involving the symplectic invariants:
\begin{equation}\label{PosCondInequalities}
\det J_{AB} - \Delta + 1 \geq 0, \quad \Delta^2 \geq 4 \det J_{AB}, \quad \sigma_{AB} > 0.
\end{equation}

We are now in a position to state and prove a necessary and sufficient condition for entanglement in terms of the symmetric overlap.\\

\begin{tcolorbox}[colframe=gray, colback=white, breakable]
\begin{theorem}
Let $A$ and $B$ be  two-mode single subsystem $AB$   prepared  in a Gaussian state $\hat \rho_{AB}$. Let's assume $A$ and $B$ are mutually correlated. The modes $A$ and $B$ are entangled with each other if and only if 
\begin{align} 
 \quad \mathcal{D}^{sym} >\mathcal{D}_{c},
\end{align}
where the state-dependent threshold  critical value $\mathcal{D}_{c}$ is given by
\begin{align}\label{eq:D_c}&\mathcal{D}_{c}:=\\& \nonumber \frac12 \left( \frac{\det J_{AB} - \det J_A}{\det J_B-1} + \frac{\det J_{AB} - \det J_B}{\det J_A-1}\right) -1 \, .\end{align}
\end{theorem}
\end{tcolorbox}

\begin{proof}
Note first that Eq.~\eqref{PosCondInequalities} is satisfied for any Gaussian state $\hat{\rho}_{AB}$.

Under partial transposition on one subsystem, among the symplectic invariants $\det J_{AB}$, $\det J_A$, $\det J_B$, and $\det J_C$, only $\det J_C$ is affected, changing its sign. Therefore, the partially transposed symplectic eigenvalues read
\begin{equation}\label{SympEigenValuesTwoModes}
\tilde{\nu}_{\pm}^{2} = \frac{\tilde{\Delta} \pm \sqrt{\tilde{\Delta}^2 - 4 \det J_{AB}}}{2},
\end{equation}
where
\begin{equation}\label{DeltaTilde}
\tilde{\Delta} = \det J_{A} + \det J_B - 2\det J_C.
\end{equation}

We further examine how the partial transposition modifies the condition $\nu_\pm \geq 1$, or equivalently, Eq.~\eqref{PosCondInequalities}. The third condition in Eq.~\eqref{PosCondInequalities} remains valid, as the partial transposition preserves the positivity of the covariance matrix. The first condition is now translated into
\begin{align}\label{SufficientConditionEnt2M}
\det J_{AB} - \tilde{\Delta} + 1 \geq 0,
\end{align}
while the second condition becomes $\tilde{\Delta}^2 \geq 4 \det J_{AB}$.

Using Eq.~\eqref{DeltaTilde}, the violation of Eq.~\eqref{SufficientConditionEnt2M}, which serves as a sufficient condition for entanglement in terms of $\det J_C$, can be written as
\begin{align}
-2 \det J_C>\det J_{AB} + 1 - \det J_A - \det J_B .
\end{align}
Substituting $\mathcal{D}^{sym}$ into the left-hand side of this inequality, and after some algebra, we obtain
\begin{align}\label{Dceformula}
 \mathcal{D}^{sym}>\mathcal{D}_{c},
\end{align}
is sufficient for entanglement, where
\begin{align}
\mathcal{D}_{c} := \frac{1}{2} \left( \frac{\det J_{AB} - \det J_A}{\det J_B - 1} + \frac{\det J_{AB} - \det J_B}{\det J_A - 1} \right) - 1.
\end{align}

Next, we show that Eq.~\eqref{Dceformula} also provides a necessary condition for entanglement between modes $A$ and $B$ in a Gaussian state---i.e, $\mathcal{D}^{sym} \leq \mathcal{D}_{c}$ implies no entanglement.
That $\mathcal{D}^{sym} \leq \mathcal{D}_{c}$ implies no entanglement, follows from the the following two cases:
\begin{itemize}
\item If $\mathcal{D}^{sym} \leq 0$, then $\det J_C \geq 0$, and by the Gaussian separability lemma the state $\hat{\rho}_{AB}$ is separable.
\item If $\mathcal{D}^{sym} > 0$, then $\det J_C < 0$, and $\tilde{\Delta} > \Delta$ holds, implying that $\tilde{\Delta}^2 \geq \Delta^2 \geq 4\det J_{AB}$. Combined with the condition  $\mathcal{D}^{sym} \leq \mathcal{D}_{c}$, this shows that the partially transposed state $\hat{\rho}^{T_A}_{AB}$ represents a Gaussian state satisfying the separability condition since $-\det J_C \geq 0$. Consequently, $\hat{\rho}^{T_A}_{AB}$ is separable, and because separability is preserved under partial transposition, the original Gaussian state $\hat{\rho}_{AB}$ is also separable.
\end{itemize}
Therefore, $\mathcal{D}^{sym} > \mathcal{D}_{c}$ yields a necessary and sufficient condition for entanglement for any Gaussian state $\hat{\rho}_{AB} $. 
\end{proof}
\begin{widetext}

\begin{tcolorbox}[colframe=gray, colback=yellow!2, breakable]
\textbf{Example:  2-mode squeezed vacuum combined with a third mode through a beam splitter}\\

This is a continuation of  the example written in the previous section, in which we consider three oscillators $ABC$ in a state prepared by performing  a beam splitter transformation between $B$ and $C$ to a two-mode squeezed vacuum between for $A$ and $B$.

The purification partner of $A$ was worked out in the previous section:
\[
\Gamma_{A_p} = \mathrm{span}\left[ i (\cos \theta\,  \gamma_B, - \,  \sin\theta \, \gamma_C)\, , - i (\cos \theta \, \gamma_B^{*}, - \,  \sin\theta \, \gamma_C^{*}) \right].
\]
i.e., the partner of $A$ is distributed between $B$ and $C$.  

Entanglement between $A$ and $B$ is characterized by the restriction of the partially transposed complex structure $J^{T_A}$ to the subspace $\Gamma_{A}\oplus \Gamma_{B}$, given by
\begin{align}
J^{T_A}_{AB} = \Pi_{AB}  J^{T_A} \, \Pi_{AB},
\end{align}
where $\Pi_{AB} $ refers to the projector into the subspace $\Gamma_{A}\oplus \Gamma_{B}$. From the restricted two-mode complex structure we can obtain the smallest symplectic eigenvalue and compute Logarithmic Negativity. This is represented in Fig. \ref{Fig:En-OV-HO}.\\

On the other hand, we  compute $\mathcal{D}^{sym}$ and the threshold critical value $\mathcal{D}_{c}$ from  \eqref{OverlapFormulaDeterminants} and \eqref{eq:D_c}. These can be obtained analytically and are given by
\begin{align}
\mathcal{D}^{sym} &= \cos ^2(\theta ) \left(\frac{\sinh ^2(2 r)}{\left(\cos (2 \theta ) \sinh ^2r+\cosh ^2 r\right)^2-1}+1\right)\\
\mathcal{D}_{c} &= -\cos ^2\theta-\frac{4 \cosh ^2r}{2 \cos (2 \theta ) \sinh ^2r+\cosh (2 r)+3}
\end{align}

In Fig. \ref{Fig:En-OV-HO}, we see that $\mathcal{D}^{sym} - \mathcal{D}_{c}$ is bigger than zero and oscillates with $\theta$ in the same way   Logarithmic Negativity does. Note that when $\theta = (n+1/2)\pi$ for $n\in \mathbb{N}$, mode $B$ is in a pure state. Therefore, $A$ and $B$ are not correlated and the necessary and sufficient condition for entanglement, $\mathcal{D}^{sym}>\mathcal{D}_c$, does not apply in this case. These values of $\theta$  are indicated by the purple vertical lines. For any other value of $\theta$ the figure shows that  non zero $\mathcal{D}^{sym}-\mathcal{D}_c$ corresponds to non-zero entanglement, and that $\mathcal{D}^{sym}-\mathcal{D}_c$ follows the same trend as Logarithmic Negativity. 

\begin{figure}[H]
    \centering
    \includegraphics[width = 0.7\textwidth]{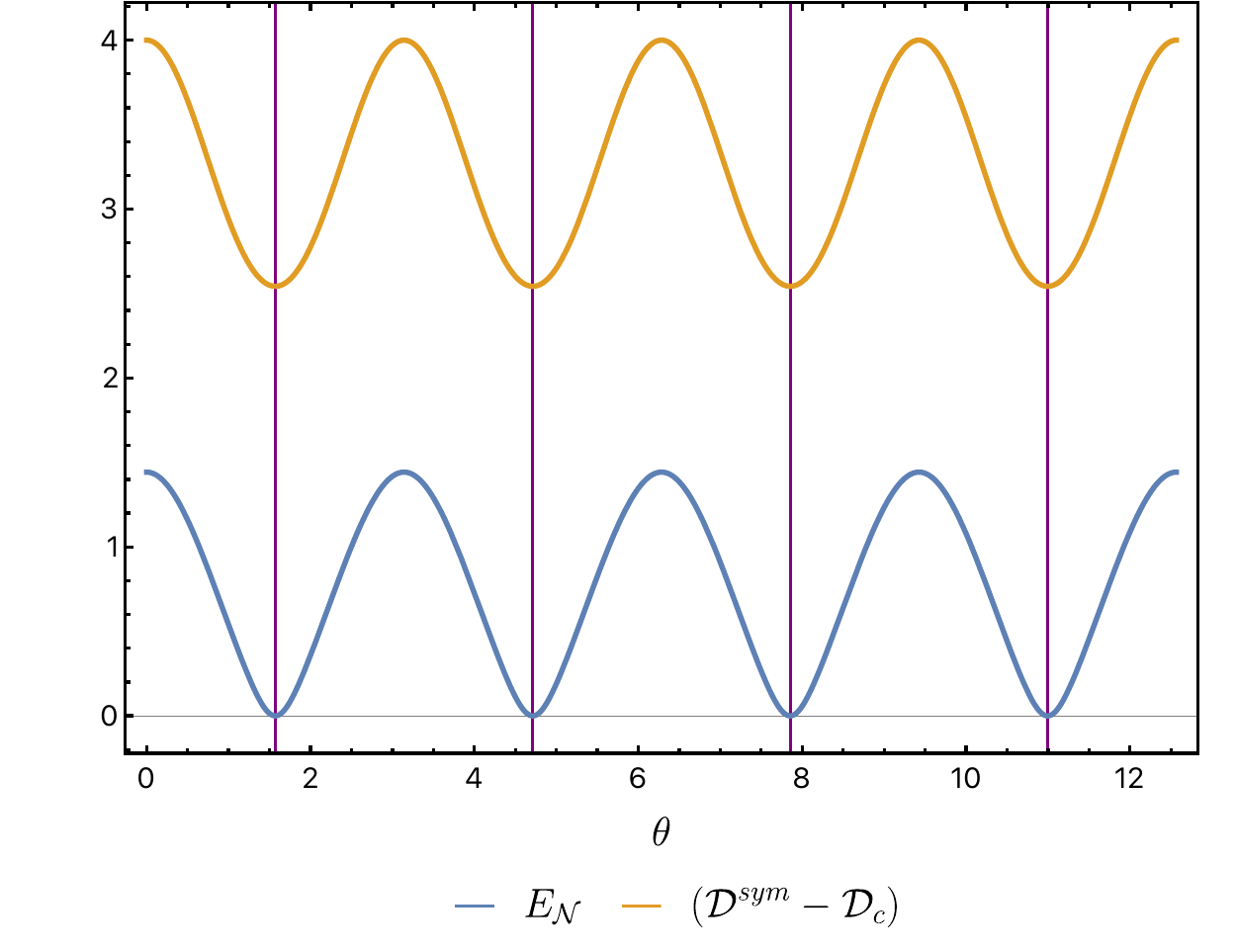}
    \caption{Entanglement between $A$ and $B$ as a function of the Beam splitter parameter $\theta$ for a squeezer parameter $r=0.5$. The blue and orange lines describes the Logarithmic Negativity $E_{\mathcal{N}}$ and the difference $\mathcal{D}^{sym} - \mathcal{D}_{c}$, respectively.
    The purple vertical lines represent the values for $\theta$ where $A$ and $B$ are not correlated and the state is separable. }
    \label{Fig:En-OV-HO}
\end{figure}

\end{tcolorbox}
\end{widetext}

\section{Extension to quantum field theory}

Classically, a field on a continuous spacetime constitutes a physical system with infinitely many degrees of freedom. The corresponding quantum field therefore contains infinitely many quantum modes. For bosonic fields, each individual mode defines a continuous-variable single-mode subsystem of the type discussed in the previous sections. Consequently, for any two modes of a quantum field prepared in a Gaussian state, all of the results developed so far in this article apply directly. The only additional task is the computation of the reduced state of the two-mode subsystem $(A,B)$ from the state  the quantum field theory is prepared in. Although the procedure is analogous to the finite-dimensional case, it involves subtleties that are intrinsic to quantum field theory.

In this section, we summarize the relevant tools, following \cite{QFTpartner} (see also \cite{Ribes-Metidieri:2025nfw}), and discuss the rather straightforward extension of the symmetric overlap to QFT. In the next section, we  illustrate the computation of the symmetric overlap and its relation to entanglement through a family of representative examples.

Consider a real scalar field  obeying the Klein–Gordon equation in any globally hyperbolic $3+1$-dimensional spacetime (generalization to other dimensions and to other bosonic fields is straightforward).

Recall that in quantum field theory the object $\hat{\phi}(x)$ does not define an operator in any reasonable way, and must be interpreted in a distributional manner. Well-defined operators can be constructed by integrating (smearing) $\hat{\phi}(x)$ against suitably chosen real functions:
\begin{equation} \label{covopt}
    \hat{\Phi}(F) = \int \text{d}V \, F(x)\, \hat{\phi}(x),
\end{equation}
where $\text{d}V = \text{d}^4x\,\sqrt{g}$ is the spacetime volume element.

For the construction of partner modes, it is convenient to work in the canonical picture. This requires the introduction of a foliation of spacetime into a one-parameter family of spatial Cauchy hypersurfaces $\Sigma_t$. For any fixed value of $t$, we can define
\begin{align}
    \hat{\Phi}(\vec{x}) &:= \phi(x)\big|_{t}, \nonumber \\
    \hat{\Pi}(\vec{x}) &:= \left(\sqrt{h}\, n^a \nabla_a \phi(x)\right)\bigg|_{t}, \nonumber
\end{align}
where $n^a$ is the future-oriented unit normal to $\Sigma_t$, $\nabla_a$ any derivative operator, and $\sqrt{h}$ is the determinant of the metric induced on $\Sigma_t$.

Smearing $\hat{\Phi}(\vec{x})$ and $\hat{\Pi}(\vec{x})$ with functions defined on $\Sigma_t$ one constructs operators that are linear in the field and in the momentum, respectively:
\begin{equation}\label{smearedphi}
    \hat{\Phi}(f) := \int_{\Sigma_t} \text{d}^3x\, f(\vec{x})\, \hat{\Phi}(\vec{x}), 
    \ \ 
    \hat{\Pi}(g) := \int_{\Sigma_t} \text{d}^3x\, g(\vec{x})\, \hat{\Pi}(\vec{x}).
\end{equation}
(The function $f(\vec{x})$ is assumed to have density weight one, so that it can be integrated against a scalar field.)
General linear observables are linear combinations of operators of this type.

We can organize all linear operators in the same way as for finite-dimensional systems in Section~\ref{sec:preliminaries}, namely by introducing a one-to-one correspondence between  such operators and element of the classical phase space. This is done as follows.

Let $\Gamma$ be the classical phase space,\footnote{$\Gamma$ is an infinite-dimensional vector space. Vectors in $\Gamma$ are made of pairs of real smooth functions of compact support defined in $\Sigma_t$, $\gamma=(g(\vec{x}),f(\vec{x}))$, having density weights zero and one, respectively—consistent with $g(\vec{x})$ and $f(\vec{x})$ describing the field and momentum components.} equipped with a symplectic structure $\Omega$. Let $\Gamma_{\mathbb{C}}$ denote the complexified phase space; the domain of $\Omega$ can be extended to $\Gamma_{\mathbb{C}}$  by linearity.

For any $\gamma=(g(\vec{x}),f(\vec{x}))$ and $\gamma'=(g'(\vec{x}),f'(\vec{x}))$ in $\Gamma_{\mathbb{C}}$, their complexified symplectic product—also known  in this context as the Klein–Gordon product—is defined as 
\begin{align}\label{sympprodQFT}
\langle \gamma, \gamma' \rangle 
&\coloneqq \frac{1}{\ii}\,\Omega(\gamma^{*},\gamma') \\
&= \frac{1}{\ii}\int_{\Sigma_t} \text{d}^3x \,\bigl( f^{*}(\vec{x})\, g'(\vec{x}) - g^{*}(\vec{x})\, f'(\vec{x}) \bigr). \nonumber
\end{align}

To each vector in $\Gamma_{\mathbb{C}}$ we associate a linear operator via
\begin{equation} \label{FieldOpdef}
    \gamma \in \Gamma_{\mathbb{C}} \longrightarrow 
    \hat{O}_{\gamma} = \ii\,\langle \gamma, \hat{\boldsymbol{R}} \rangle,
\end{equation}
where $\hat{\boldsymbol{R}} := (\hat{\Phi}(\vec{x}),\hat{\Pi}(\vec{x}))$.  
In this way, the complex vector $\gamma = (g(\vec{x}),f(\vec{x}))$ defines the operator
\begin{equation} \label{mapgtop}
    \hat{O}_{\gamma} 
    = \int_{\Sigma_t} \text{d}^3x\, \bigl( f^{*}\,\hat{\Phi} - g^{*}\,\hat{\Pi} \bigr).
\end{equation}
The commutation relations between any two such operators are
\begin{equation} \label{comm}
    [\hat{O}_{\gamma}, \hat{O}_{\gamma'}^{\dagger}] 
    = \langle \gamma, \gamma' \rangle,
\end{equation}
which follow directly from the equal-time canonical commutation relations between the field and its momentum.

Finally, as in mechanical systems, subsystems are in one-to-one correspondence with \emph{symplectic subspaces} of the phase space. Using the relation $\gamma \mapsto \hat{O}_{\gamma}$, a symplectic subspace $\Gamma_A$ defines a subalgebra of quantum observables—generated by products of operators $\hat{O}_{\gamma}$ in the standard way—with $\gamma \in \Gamma_A$. In the algebraic approach to quantum field theory, this subalgebra defines a quantum subsystem. This establishes the relation between classical subsystems and their quantum counterparts.

If $\Gamma_A$ has (finite) dimension $2N_A$ with $N_A \in \mathbb{N}$, the associated quantum subsystem is said to be an $N_A$-mode system, and the associated algebra is isomorphic to that of $N_A$ harmonic oscillators.

\subsection{Gaussian states and complex structures in QFT}
As discussed in previous sections, a Gaussian state with density operator $\hat{\rho}$ is fully characterized by a covector $\mu_a$ and a twice-covariant tensor $\sigma_{ab}$ in $\Gamma_{\mathbb{C}}$, encoding the first and second moments, respectively. In QFT, the covector $\mu_a$ has the form
\begin{equation}
    \mu_a(\vec{x}) = \bigl( \langle \hat{\Pi}(\vec{x}) \rangle,\; \langle \hat{\Phi}(\vec{x}) \rangle \bigr),
\end{equation}
where we are adopting the notation $\operatorname{Tr}\!\,[ \hat{\rho}\,\hat{O}] \equiv \langle \hat{O} \rangle$. The covariance matrix $\sigma_{ab}$ is given by the bidistribution
\begin{equation}
\sigma(\vec{x},\vec{x}')
=
\begin{pmatrix}
\langle \{ \hat{\bar\Pi}(\vec{x}), \hat{\bar\Pi}(\vec{x}') \} \rangle
&
-\langle \{ \hat{\bar\Pi}(\vec{x}), \hat{\bar\Phi}(\vec{x}') \} \rangle
\\[6pt]
-\langle \{ \hat{\bar\Phi}(\vec{x}), \hat{\bar\Pi}(\vec{x}') \} \rangle
&
\langle \{ \hat{\bar\Phi}(\vec{x}), \hat{\bar\Phi}(\vec{x}') \} \rangle
\end{pmatrix},
\end{equation}
where $\hat{\bar\Phi} := \hat{\Phi} - \langle \hat{\Phi} \rangle$ denotes the centered field operator (and similarly for $\hat{\bar\Pi}$).

Combining $\sigma$ and $\Omega$, one obtains a linear map on $\Gamma_{\mathbb{C}}$ (the analog of $J^{a}{}_{b}$ defined in~\eqref{J_definition}):
\[
J(\vec{x},\vec{x}')
=
-\!\int \text{d}^3x'' \,\Omega(\vec{x},\vec{x}'') \,\sigma(\vec{x}'',\vec{x}').
\]
Here $\Omega(\vec{x},\vec{x}')$ denotes the inverse symplectic structure, given by
\[
\Omega(\vec{x},\vec{x}') =
\begin{pmatrix}
0 & 1 \\
-1 & 0
\end{pmatrix}
\delta(\vec{x}-\vec{x}'),
\]
(which is invertible in the Cauchy completion of $\Gamma$ using the inner product defined by $\sigma$.)

A Gaussian state is pure if and only if $J$ satisfies $J^{2} = -\mathbb{I}$, i.e.\ $J$ defines a \emph{complex structure}. For mixed states, $J$ satisfies $J^{2} < -\mathbb{I}$ and defines a \emph{restricted complex structure}. 

Let $\Gamma_A$ be a single-mode subsystem, and let $\{\gamma_A^{(1)},\gamma_A^{(2)}\}$ be a Darboux basis in $\Gamma_A$. Then the restriction of the symplectic structure to $\Gamma_A$ is
\begin{equation}
\Omega_A = 
\begin{pmatrix}
\Omega(\gamma_A^{(1)},\gamma_A^{(1)}) & \Omega(\gamma_A^{(1)},\gamma_A^{(2)}) \\
\Omega(\gamma_A^{(2)},\gamma_A^{(1)}) & \Omega(\gamma_A^{(2)},\gamma_A^{(2)})
\end{pmatrix}
=
\begin{pmatrix}
0 & -1 \\
1 & \phantom{-}0
\end{pmatrix}.
\end{equation}

Given a Gaussian state $\hat{\rho}$ with first moment $\mu(\vec{x})$ and covariance $\sigma(\vec{x},\vec{x}')$, the reduced state on subsystem $A$ is another Gaussian state with first moment
\[
\mu_A = \bigl( \mu(\gamma_A^{(1)}),\; \mu(\gamma_A^{(2)}) \bigr)
\]
and covariance matrix
\begin{eqnarray}
\sigma_A &=
\begin{pmatrix}
\sigma(\gamma_A^{(1)},\gamma_A^{(1)}) 
&
\sigma(\gamma_A^{(1)},\gamma_A^{(2)}) 
\\[6pt]
\sigma(\gamma_A^{(2)},\gamma_A^{(1)}) 
&
\sigma(\gamma_A^{(2)},\gamma_A^{(2)})
\end{pmatrix}\nonumber \\ & =-\ii \begin{pmatrix}
       \langle \gamma^{(1)*}_A, J\gamma^{(1)}_A\rangle &  \langle \gamma^{(1)*}_A, J\gamma^{(2)}_A\rangle\\ \langle \gamma^{(2)*}_A, J\gamma^{(1)}_A\rangle & \langle \gamma^{(2)*}_A, J\gamma^{(2)}_A\rangle\end{pmatrix},
\end{eqnarray}
where we have used that $\sigma(\gamma,\gamma')$ can be written as $-\ii \langle \gamma^*,J\gamma'\rangle$.

Hence, the calculation of the reduced state of a Gaussian state in quantum field theory reduces to computing the above quantities. (See \cite{Ribes-Metidieri:2025nfw,QFTpartner} for further details.) In the next section, we present a  explicit examples illustrating these constructions.

\subsection{Partner mode and overlap measure}
When the field theory is prepared in a pure Gaussian state, the construction of the purification partner of a given single-mode subsystem is parallel to the procedure explained in Sec.~\ref{sec:partner_review} (see \cite{agullo_correlation_2025} for further details and extensions to mixed states). Namely, if $\Gamma_A \subset \Gamma_{\mathbb{C}}$ is the symplectic subspace characterizing the mode $A$, the partner mode corresponds to
\begin{equation}\label{singl_corr_partner_formula2}
\Gamma_{A_p} = \Pi_A^{\perp}\!\left(J\,\Gamma_A\right) \, .
\end{equation}
If $\{\gamma_A,\gamma_A^*\}$ is a symplectic-orthonormal basis in $A$, an orthonormal basis in $\Gamma_{A_p}$ is $\{\gamma_{A_p},\gamma_{A_p}^*\}$, where 

\begin{equation}\label{partner}
\gamma_{A_p} = \frac{1}{\sqrt{\det J_A-1}}\Pi_A^{\perp} (J \gamma_A^{*}) \,  .
\end{equation}

Given two single-mode subsystems $A$ and $B$, their symmetric overlap is defined in the same way as in Eq.~\eqref{eqSymOverlap}, namely
\begin{equation}\label{eqSymOverlap2}
\mathcal{D}^{\mathrm{sym}} =
\langle \Pi_{A_p}(\gamma_B), \Pi_{A_p}(\gamma_B) \rangle
+
\langle \Pi_{B_p}(\gamma_A), \Pi_{B_p}(\gamma_A) \rangle \, ,
\end{equation}
where $\gamma_A$ and $\gamma_B$ are any unit-symplectic-norm vectors in $\Gamma_A$ and $\Gamma_B$, respectively.

Using the same steps as in the previous section, this quantity can be rewritten in terms of the restricted complex structure matrix $J_{AB}$ of the two-mode system $(A,B)$ as
\begin{equation}\label{OverlapFormulaDeterminants2QFT}
\mathcal{D}^{\mathrm{sym}} =
\left(
\frac{1}{\det J_B - 1}
+
\frac{1}{\det J_A - 1}
\right)
\bigl(-\det J_C\bigr)\, ,
\end{equation}
where, 
given complex orthonormal basis  in $\Gamma_A$ and $\Gamma_B$---denoted by
$\{\gamma_A,\gamma_A^*\}$ and
$\{\gamma_B,\gamma_B^*\}$, respectively---the  restricted complex structure matrix of the $(A,B)$ system is
\begin{widetext}
\begin{equation}\label{TwoModesNonHerm_CovMatrixQFT}
J_{AB}
=
\begin{pmatrix}
J_A & J_C\\
J_C^{\top} & J_B
\end{pmatrix}
= \begin{pmatrix}
\langle \gamma_A, J\gamma_A\rangle
&
\langle \gamma_A, J\gamma_A^{*}\rangle
&
\langle \gamma_A, J\gamma_B\rangle
&
\langle \gamma_A, J\gamma_B^{*}\rangle
\\
-\langle \gamma_A^*, J\gamma_A\rangle
&
-\langle \gamma_A^*, J\gamma_A^{*}\rangle
&
-\langle \gamma_A^*, J\gamma_B\rangle
&
-\langle \gamma_A^*, J\gamma_B^{*}\rangle\\\langle \gamma_B, J\gamma_A\rangle
&
\langle \gamma_B, J\gamma_A^{*}\rangle
&
\langle \gamma_B, J\gamma_B\rangle
&
\langle \gamma_B, J\gamma_B^{*}\rangle\\-\langle \gamma_B^*, J\gamma_A\rangle
&
-\langle \gamma_B^*, J\gamma_A^{*}\rangle
&
-\langle \gamma_B^*, J\gamma_B\rangle
&
-\langle \gamma_B^*, J\gamma_B^{*}\rangle
\end{pmatrix}.
\end{equation} 
\end{widetext}

From Eqs.~\eqref{TwoModesNonHerm_CovMatrixQFT} and \eqref{OverlapFormulaDeterminants2QFT} one can check that the computation of $\mathcal{D}^{\mathrm{sym}}$ reduces to evaluating the ten independent symplectic products appearing in Eq.~\eqref{TwoModesNonHerm_CovMatrixQFT}. 

As in quantum mechanics, in QFT it remains true that $\mathcal{D}^{\mathrm{sym}}$ depends only on the reduced state of the two-mode system $(A,B)$. This allows one to extend the applicability of Eq.~\eqref{OverlapFormulaDeterminants2QFT} to situations in which the total state of the field is mixed.

The  results described in subsection \ref{subsec:NecessarySufficientOverlap} remain true in QFT---and the proofs are identical. Namely, two single-mode subsystems are entangled if and only if $\mathcal{D}^{\mathrm{sym}}>\mathcal{D}_c$, where 
\begin{align}\label{DcQFT}
\mathcal{D}_{c}=\frac12 \left( \frac{\det J_{AB} - \det J_A}{\det J_B-1} + \frac{\det J_{AB} - \det J_B}{\det J_A-1}\right) -1 \, .
\end{align}

\section{Example: Ball-Shell entanglement}
This section illustrates the calculation of the symmetric overlap between two single modes in the context of a massive scalar field theory in Minkowski spacetime prepared in the vacuum state. This state is a Gaussian state with vanishing first moments, $\mu_M(\vec{x})=(0,0)$, and covariance
\begin{equation}
\sigma_M(\vec{x},\vec{x}')=\int \frac{\text{d}^3k}{(2\pi)^3} \, e^{i\vec{k}\cdot(\vec{x}-\vec{x}')} \,
\begin{pmatrix}
\omega_k & 0\\
0 & \dfrac{1}{\omega_k}
\end{pmatrix},
\end{equation}
where $\omega_k=\sqrt{|\vec{k}|^2+m^2}$.  

The associated complex structure is given by
\begin{align}
J_M(\vec{x},\vec{x}')= \int \frac{\text{d}^3k}{(2\pi)^3}\, e^{i\vec{k}\cdot(\vec{x}-\vec{x}')}
\begin{pmatrix}
0 & -\dfrac{1}{\omega_k}\\
\omega_k & 0
\end{pmatrix}.
\end{align}
and satisfies
\begin{equation}
J_M^2(\vec{x},\vec{x}')=
-\begin{pmatrix}
1 & 0\\
0 & 1
\end{pmatrix}\delta^{(3)}(\vec{x}-\vec{x}'),
\end{equation}
confirming that the Minkowski vacuum is a pure Gaussian state.

\medskip

\begin{figure}
    \centering
    \includegraphics[width = 0.5\textwidth]{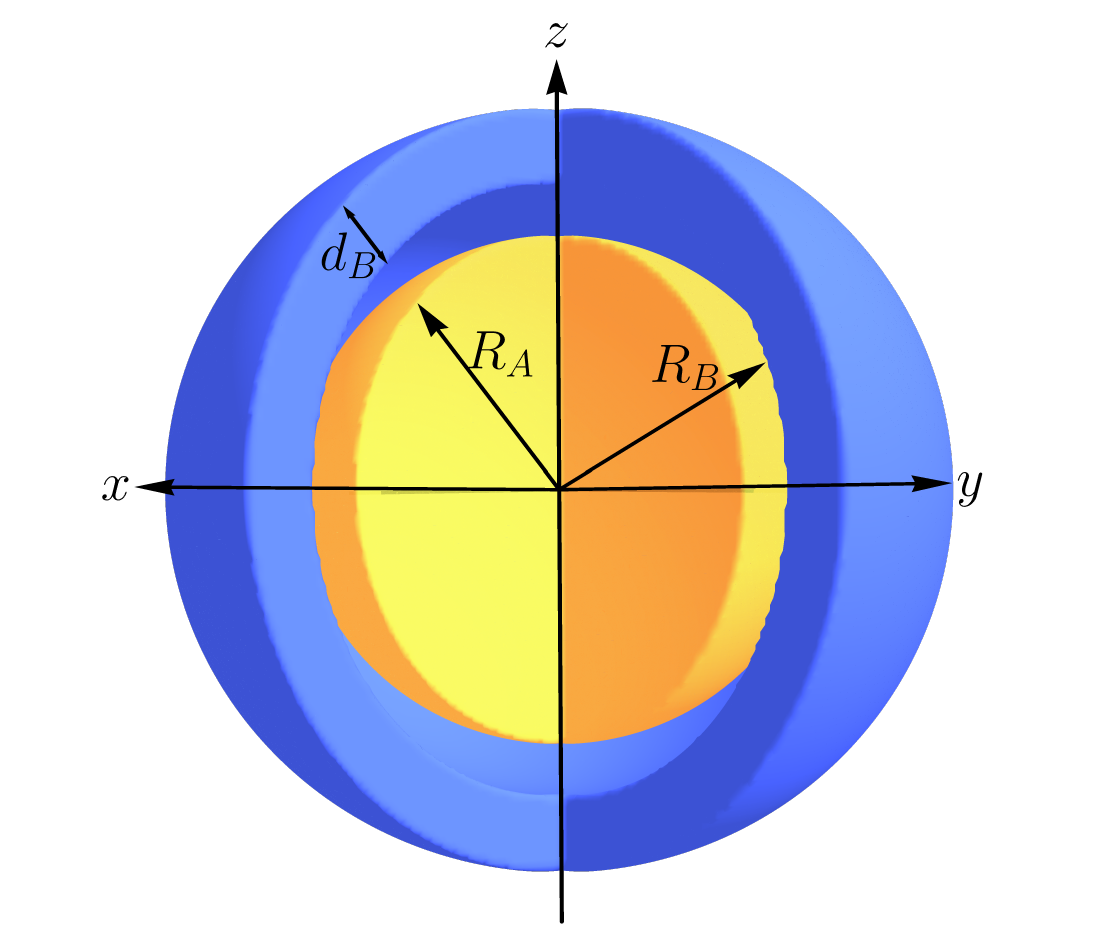}
    \caption{Illustration of the regions of support for the field modes considered in this example. Mode $A$ is supported within a ball of radius $R_A$ (light orange). Mode $B$ is supported within a spherical shell surrounding the ball, with inner radius $R_B$ and radial width $d_B$ (dark blue).}
    \label{fig:ball&shell}
\end{figure}

For subsystem $A$, we consider a mode compactly supported within a ball of radius $R_A$ [highlighted in light orange in Fig.~\ref{fig:ball&shell}].  

Such a mode can be uniquely specified by choosing two real, non-orthogonal phase-space vectors $\{\gamma_A^{(1)},\gamma_A^{(2)}\}$. These vectors can be normalized so that $\Omega(\gamma_A^{(1)},\gamma_A^{(2)})=1$; or equivalently
$\langle \gamma_A^{(1)},\gamma_A^{(2)} \rangle=\ii$. 
They span a two-dimensional symplectic subspace $\Gamma_A$, and the associated operators are canonically conjugate:
\begin{equation}
[\hat{O}_{\gamma_A^{(1)}},\hat{O}_{\gamma_A^{(2)}}]=\ii.
\end{equation}
(A complex orthonormal basis $\{\gamma_A,\gamma_A^*\}$ can be readily defined from  $\gamma_A^{(1)}$ and $\gamma_A^{(2)}$, with $\gamma_A=\frac{1}{\sqrt{2}}(\gamma_A^{(1)}-\ii\, \gamma_A^{(2)})$.)
In this example, we choose
\begin{equation}\label{vec_A}
\gamma_{A}^{(1)}(\vec{x})=
\begin{pmatrix}
0\\
f_A(\vec{x})
\end{pmatrix},
\qquad
\gamma_{A}^{(2)}(\vec{x})=
\begin{pmatrix}
-\,f_A(\vec{x})\\
0
\end{pmatrix},
\end{equation}
with
\begin{equation}\label{f_Ag_A-Ball}
f_A(\vec{x})=
K_A \cos^2\!\left(\frac{\pi}{2}\frac{|\vec{x}|}{R_A}\right)
\,\Theta(R_A-|\vec{x}|)\, ,
\end{equation}
where $\Theta(x)$ is the Heaviside step function, which ensures that $f_A$ is compactly supported within a ball of radius $R_A$. The normalization constant is
\begin{equation}\label{eq:Adelta}
K_A=\frac{2}{R_A^{3/2}}\sqrt{\frac{\pi}{2\pi^2-15}} .
\end{equation}

The function $f_A(\vec{x})$ is not smooth: while $f_A(\vec{x})$ and its first derivative are continuous, the second derivative is not. Nevertheless, the vectors $\gamma_A^{(1)}$ and $\gamma_A^{(2)}$ belong to $\Gamma_{\sigma_M}$—the Cauchy completion of the classical phase space $\Gamma$ with respect to the inner product defined by $\sigma_M$. As a consequence, the symplectic products and the action of $J_M$ on these modes are well defined. Hence, the modes considered in this example are perfectly well defined within the quantum field theory (see \cite{Ribes-Metidieri:2024vjn,ubiquitous,QFTpartner} for further details on the functional spaces to which elements of $\Gamma_{\sigma_M}$ belong).

For mode $B$, a natural choice would be a mode having the same functional form as mode $A$, but translated in space so that $A$ and $B$ do not overlap and thus define independent modes. However, such pairs of modes were shown in \cite{ubiquitous} to be unentangled, and therefore they do not provide an illustrative example for the purposes of this article.

Instead, we consider a mode $B$ supported on a spherical shell concentric with the support of $A$. This choice maximizes the spatial “proximity” of the two modes while keeping their supports disjoint, and leads to non-vanishing entanglement between $A$ and $B$.

Mode $B$ is characterized by a pair of basis vectors $\{\gamma_B^{(1)},\gamma_B^{(2)}\}$ defined by
\begin{equation}
   \gamma_{B}^{(1)} (\vec{x}) =
   \begin{pmatrix}
       0\\ f_B(\vec{x})
   \end{pmatrix},
   \qquad
   \gamma_{B}^{(2)}(\vec{x}) =
   \begin{pmatrix}
       -\, f_B(\vec{x})\\ 0
   \end{pmatrix},
\end{equation}
where the smearing function is
\begin{equation}
f_B(\vec{x}) =
K_B
\begin{cases}
\displaystyle
\sin^2\!\left(\pi\frac{|\vec{x}|-R_B}{d_B}\right),
& R_B \le |\vec{x}| \le R_B + d_B,\\[1.2ex]
0, & \text{otherwise}.
\end{cases}
\label{SFf2}
\end{equation}
Here, $K_B$ is a normalization constant fixed by the condition
\begin{equation}
\Omega(\gamma^{(1)}_{B},\gamma^{(2)}_{B})=1.
\end{equation}
The explicit expression for $K_B$ is relatively lengthy and does not provide additional  insight, and we therefore omit it.

In Eq.~\eqref{SFf2}, $R_B$ denotes the inner radius of the shell and $d_B$ its radial width (see Fig.~\ref{fig:ball&shell}).

We restrict to the case $R_A \leq R_B$, so that the supports of modes $A$ and $B$ do not overlap. This automatically guarantees that the two modes are independent.\footnote{Note that the converse is not true. For instance, there exist infinitely many distinct modes compactly supported within the ball of radius $R_A$ that are independent of $A$. Thus, it is important not to identify spatial regions with any finite set of field modes.}

The reduced complex structure  $J_{AB}$ can then be obtained by numerically evaluating the symplectic products appearing in Eq.~\eqref{TwoModesNonHerm_CovMatrixQFT}. From $J_{AB}$, all quantities of interest discussed in the previous sections can be computed.

\subsection{Entanglement}
\begin{figure}
    \centering
    \includegraphics[width=0.50\textwidth]{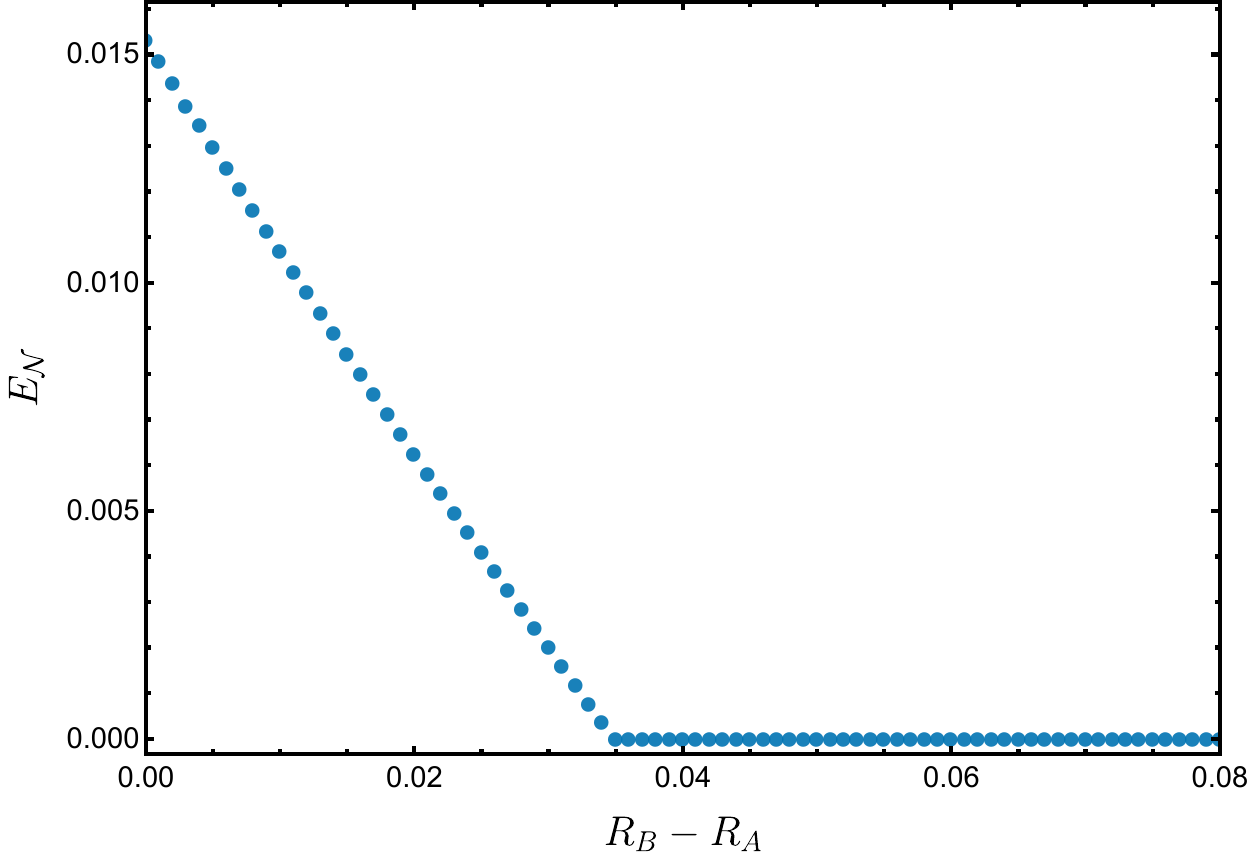}
    \caption{ Logarithmic negativity between modes $A$ and $B$, supported in a ball and a concentric spherical shell respectively, as a function of their radial separation $R_B-R_A$. The parameters are $m=0$ and $d_B=0.5$.  All lengths are measured in units of $R_A$.} 
    \label{fig:LogNeg_vs_distance} 
\end{figure}

To begin with, we compute the logarithmic negativity between modes $A$ and $B$. 
Figure~\ref{fig:LogNeg_vs_distance} shows the logarithmic negativity as a function of the radial separation $R_B-R_A$ between the modes. The plot corresponds to a massless field ($m=0$) and fixed shell thickness $d_B=0.5 $,  with all lengths measured in units of $R_A$. This choice makes the plot insensitive to the specific value of $R_A$. The same applies to all figures shown throughout the paper.

As expected, the entanglement between the modes decreases as the separation $R_B-R_A$ increases, and vanishes beyond a threshold distance. This threshold depends on the mass of the field and on the geometric parameters of the configuration. Analogous results in other spacetime dimensions were reported in \cite{ubiquitous}.

\begin{figure}
    \centering
    \includegraphics[width=0.5\textwidth]{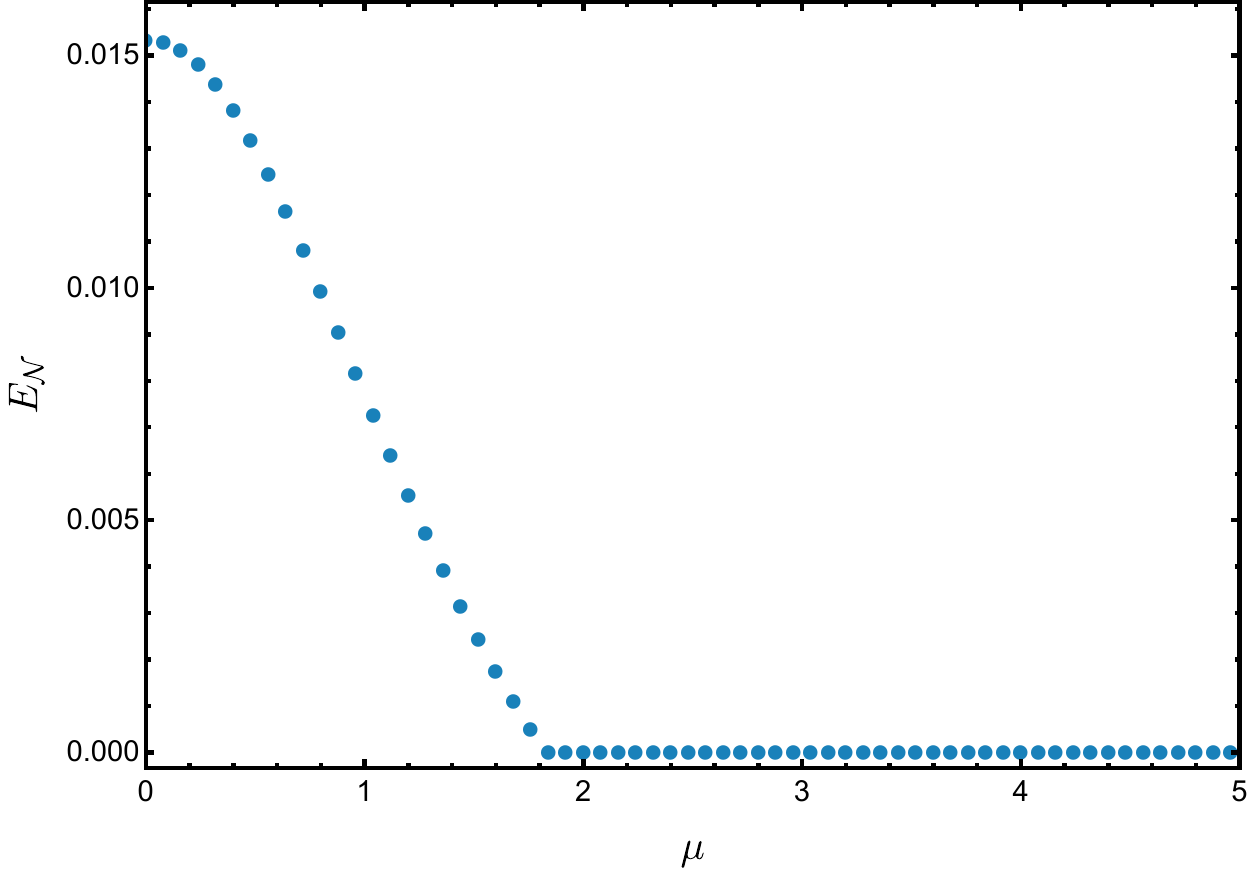}
    \caption{Logarithmic negativity as a function of the  dimensionless field mass $\mu = m R_A$. The parameters are $R_B=R_A$ (no gap between the supports of the modes) and $d_B=0.5R_A$.}
    \label{fig:LogNeg_vs_m}
\end{figure}

Figure~\ref{fig:LogNeg_vs_m} shows the logarithmic negativity as a function of the  dimensionless field mass $\mu = m R_A$. The plot shows that entanglement decreases rapidly as the mass increases. In this configuration, we set $R_A=R_B$, so that the two modes are as close as possible and therefore maximally entangled for this geometry.

\begin{figure}
    \centering
    \includegraphics[width=0.5\textwidth]{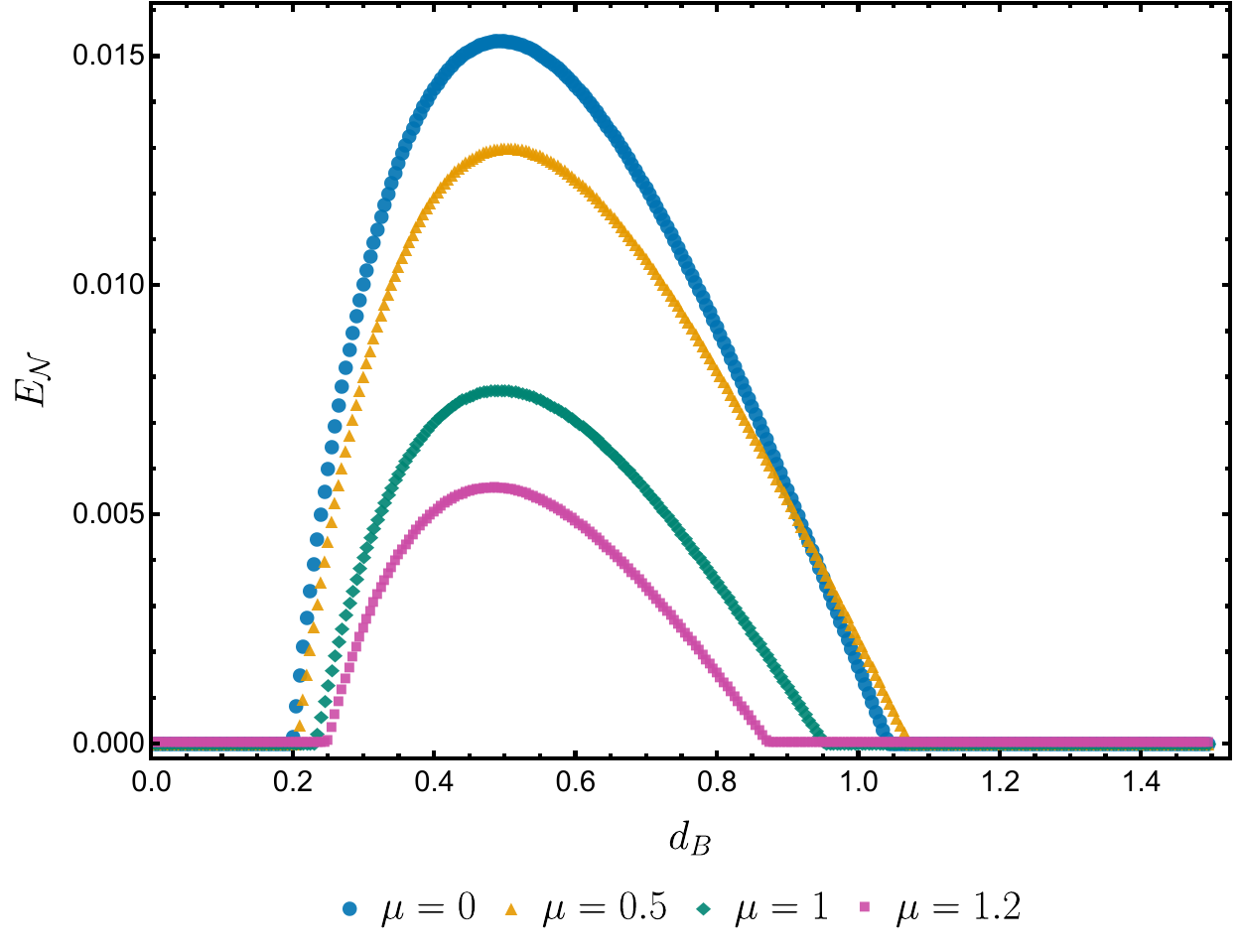}
    \caption{Logarithmic negativity as a function of the shell width $d_B$ (measured in units of $R_A$) for different values of the dimensionless field mass $\mu=m R_A$. The parameters are $R_B=R_A$. Entanglement is non-zero only for a finite range of values of $d_B$ \cite{ubiquitous}.}
    \label{fig:LogNeg_vs_dB}
\end{figure}

Finally, Fig.~\ref{fig:LogNeg_vs_dB} shows the logarithmic negativity as a function of the shell thickness $d_B$, i.e., the radial width of mode $B$.

\subsection{Partner modes}

Although not strictly necessary for computing the symmetric overlap $\mathcal{D}^{\mathrm{sym}}$ in the next subsection, for completeness we present here the explicit form of the purification partner of mode $A$, supported in the ball shown in Fig.~\ref{fig:ball&shell}.

To construct the partner mode $A_p$, it is convenient to work in a complex basis. For mode $A$, we introduce the basis $\{\gamma_A,\gamma_A^*\}$, where
\[
\gamma_A=\frac{1}{\sqrt{2}}\left(\gamma_A^{(1)}-\ii\,\gamma_A^{(2)}\right)
=\frac{1}{\sqrt{2}}\left(\ii\, f_A(\vec x),\, f_A(\vec x)\right).
\]
The basis for the partner mode $A_p$ is $\{\gamma_{A_p},\gamma_{A_p}^*\}$, with
\begin{align}\label{PartnerBasisField}
\gamma_{A_p}(\vec{x})
=
\frac{1}{\sqrt{\det J_A-1}}\,
\Pi_A^{\perp}
\int \text{d}^3 x'\, 
J_M(\vec{x},\vec{x}') \,
\gamma_A^{*}(\vec{x}') \, .
\end{align}

\begin{figure}
    \centering
\includegraphics[width=0.5\textwidth]{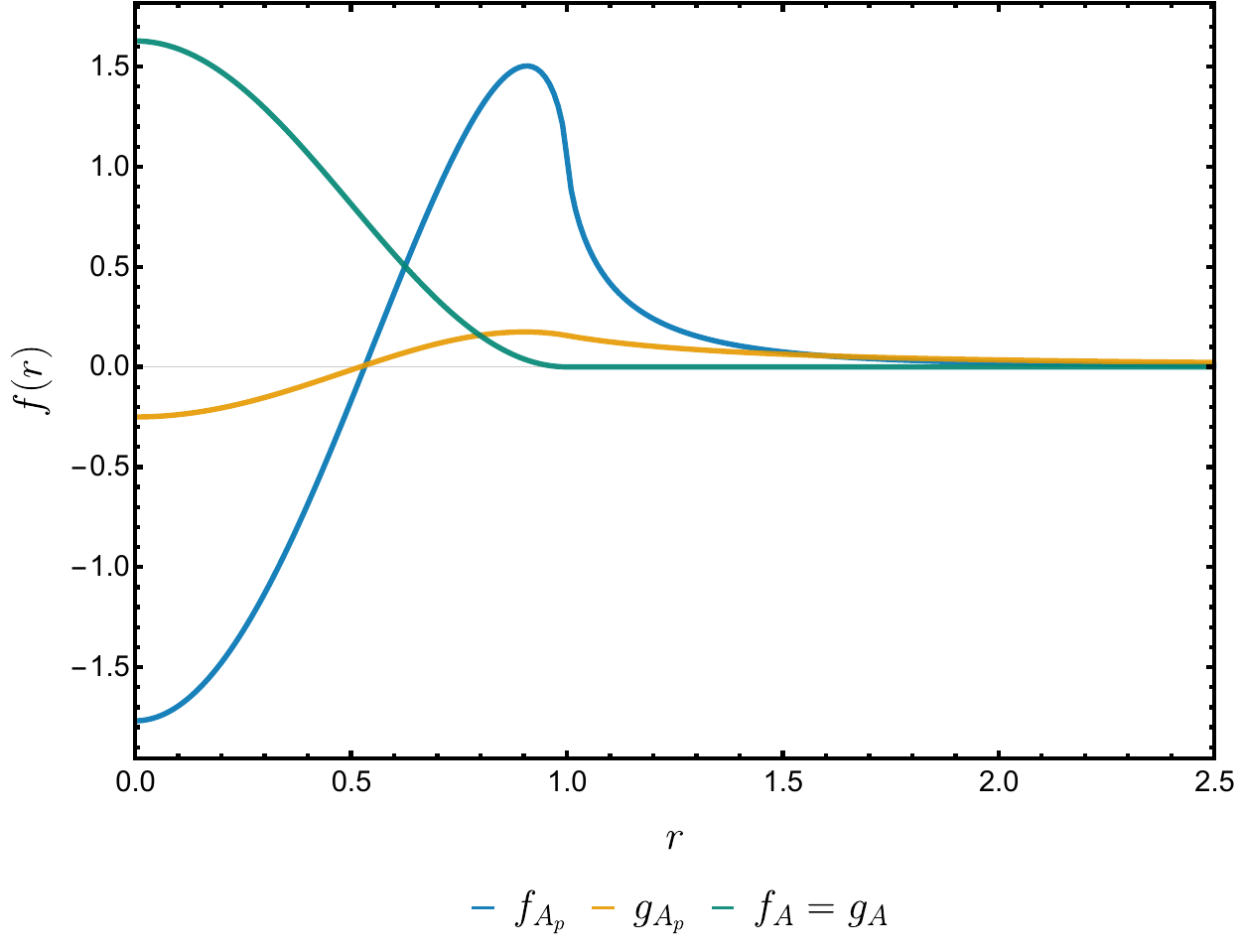}
    \caption{Shape of the spherically symmetric functions $f_{A_p}$ and $g_{A_p}$ defining the purification partner $A_p$, in terms of the dimensionless radial distance $r=\frac{|\vec{x}|}{R_A}$. The shape of the function $f_A(\vec{x})$ 
    defining mode $A$ as also shown for comparison. The figure shows that, even though  $f_A(\vec{x})$ is  compactly supported, $f_{A_p}$ and $g_{A_p}$ are not. 
    }
    \label{fig:PartnerModeA-support}
\end{figure}

\begin{figure}
    \centering  \includegraphics[width=0.5\textwidth]{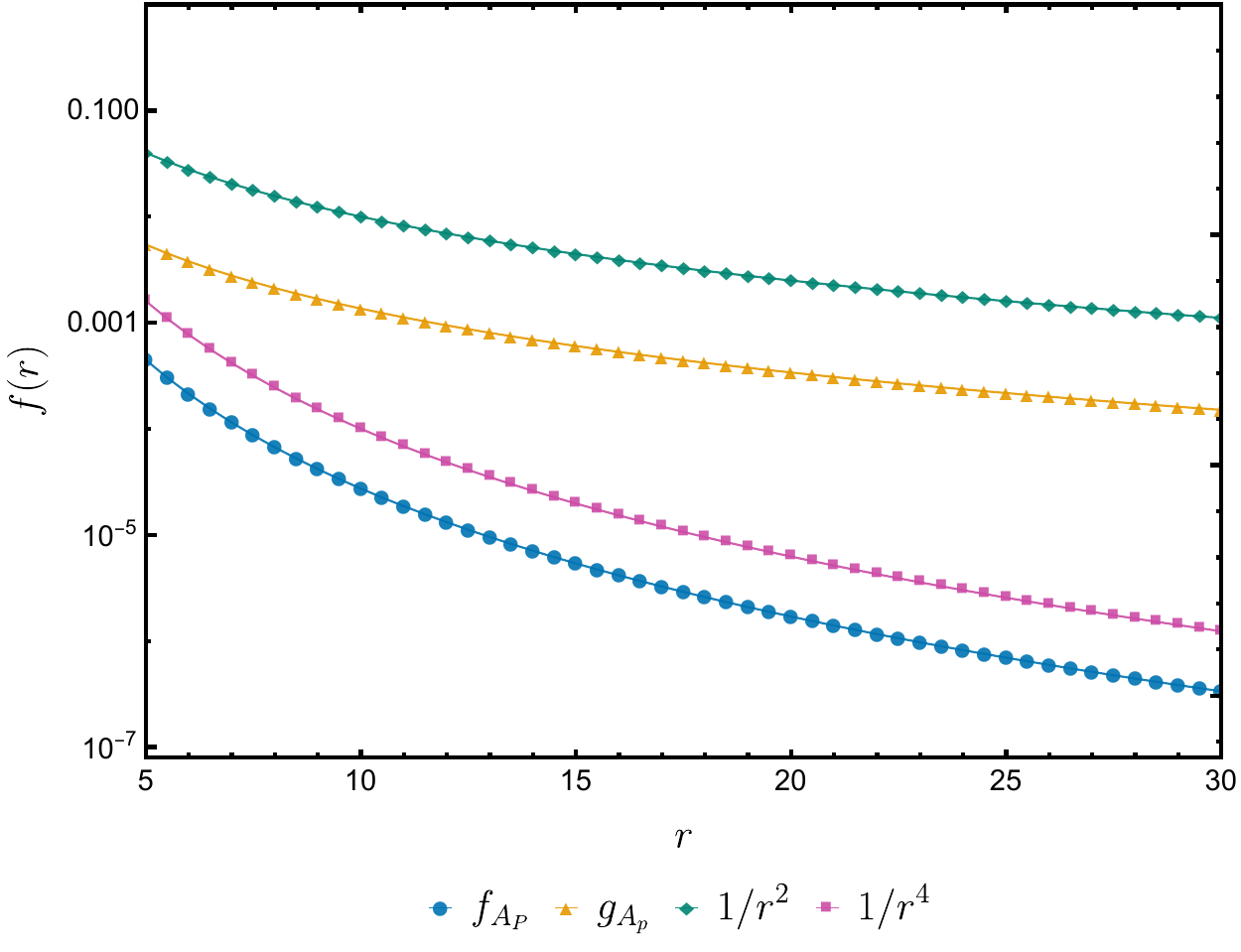}
    \caption{Asymptotic fall-off  of the functions $f_{A_p}(\vec{x})$ and $g_{A_p}(\vec{x})$.  
    This figure shows that $f_{A_p}(\vec{x})$ decays as $r^{-4}$, while $g_{A_p}(\vec{x})\sim r^{-2}$.}
    \label{fig:PartnerModeA-LargeFallOff}
\end{figure}

Since $J_M$ is real, it follows that $\gamma_{A_p}$ can be written as
\[
\gamma_{A_p}=\frac{1}{\sqrt{2}}
\left(
\ii\, g_{A_p}(\vec x),\, 
f_{A_p}(\vec x)
\right),
\]
where $f_{A_p}(\vec x)$ and $g_{A_p}(\vec x)$ are real-valued functions. 
The explicit forms of these functions are shown in Figs.~\ref{fig:PartnerModeA-support} and \ref{fig:PartnerModeA-LargeFallOff}, together with $f_A(\vec x)$ for comparison.

A key feature of the partner mode $\gamma_{A_p}$ is that it is not compactly supported, even though the original mode $A$ is compactly supported. This reflects the fact that mode $A$ is correlated and entangled with field modes arbitrarily far away, in accordance with the Reeh--Schlieder theorem \cite{ReehSchlieder}.

Figure~\ref{fig:PartnerModeA-LargeFallOff} shows the asymptotic fall-off of the partner mode, which satisfies
\[
f_{A_p}(\vec x)\sim \frac{1}{r^{4}}, 
\qquad 
g_{A_p}(\vec x)\sim \frac{1}{r^{2}},
\qquad 
r=\frac{|\vec{x}|}{R_A}.
\]
As explained in \cite{QFTpartner}, this power-law decay is a universal feature of purification partners of compactly supported modes in the Minkowski vacuum.

\subsection{Symmetric Overlap}
As emphasized above, the symmetric overlap $\mathcal{D}^{\mathrm{sym}}$ and the threshold value $\mathcal{D}_c$ can be computed directly from the restricted complex structure matrix $J_{AB}$, which we have evaluated numerically for the example considered in this section.

\begin{figure}
    \centering
    \includegraphics[width=0.5\textwidth]{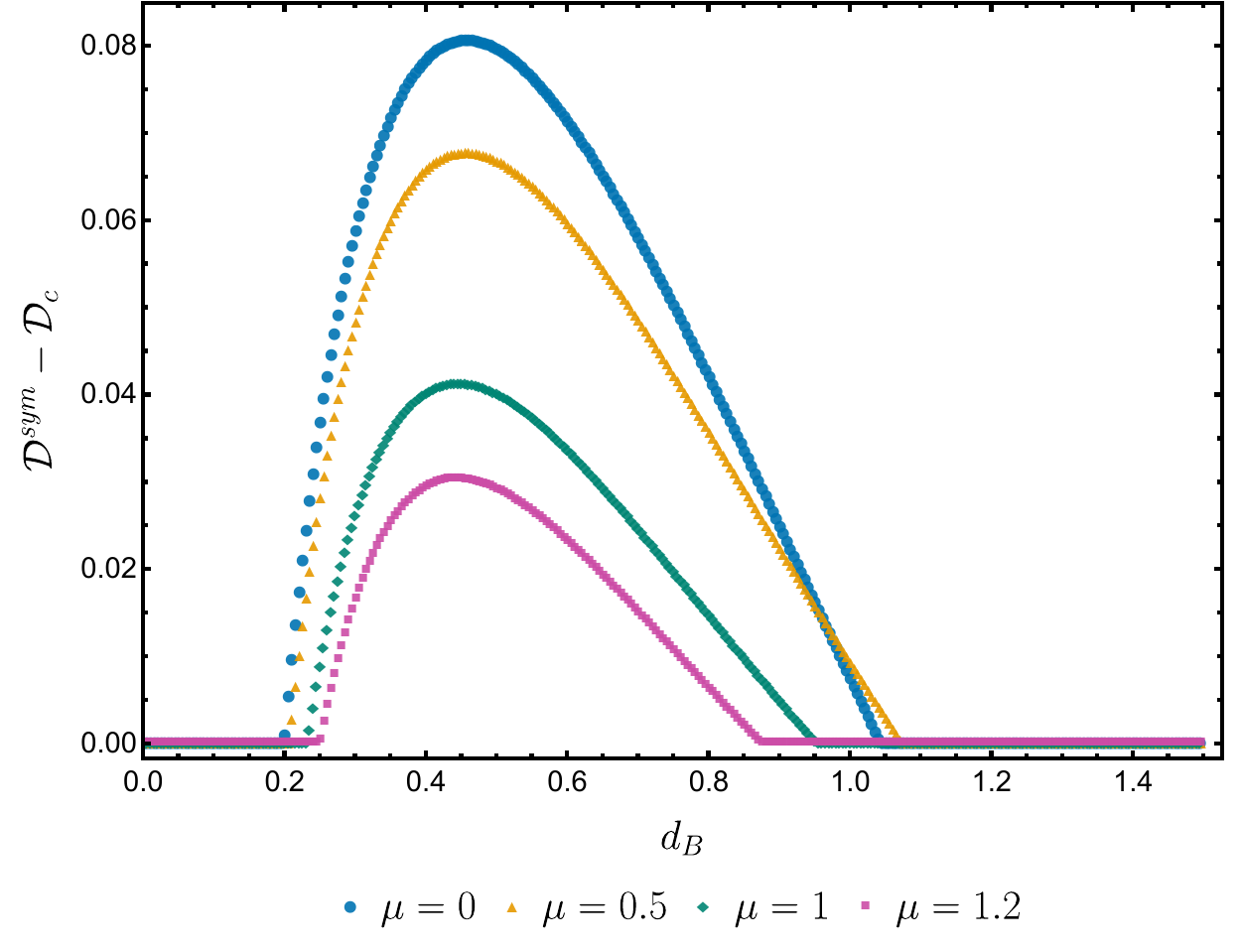}
    \caption{$\mathcal{D}^{\mathrm{sym}}-\mathcal{D}_c$ for the ball--shell configuration as a function of the shell’s radial width $d_B$ in units of $R_A$---for different values of the dimensionless field mass $\mu=m R_A$. The quantity $\mathcal{D}^{\mathrm{sym}}-\mathcal{D}_c$ is positive precisely when the two modes are entangled (compare with Fig.~\ref{fig:LogNeg_vs_dB}).}
    \label{fig:Ds-DcVSWidth}
\end{figure}

Figure~\ref{fig:Ds-DcVSWidth} shows the quantity $\mathcal{D}^{\mathrm{sym}}-\mathcal{D}_c$ for this example as a function of the radial width $d_B$ of the shell. Recall that the two modes are entangled if and only if this quantity is positive. The figure shows that $\mathcal{D}^{\mathrm{sym}}-\mathcal{D}_c$ is indeed larger than zero precisely in the same range of values of $d_B$ for which the logarithmic negativity is non-zero.

\begin{figure}
    \centering
    \includegraphics[width=0.5\textwidth]{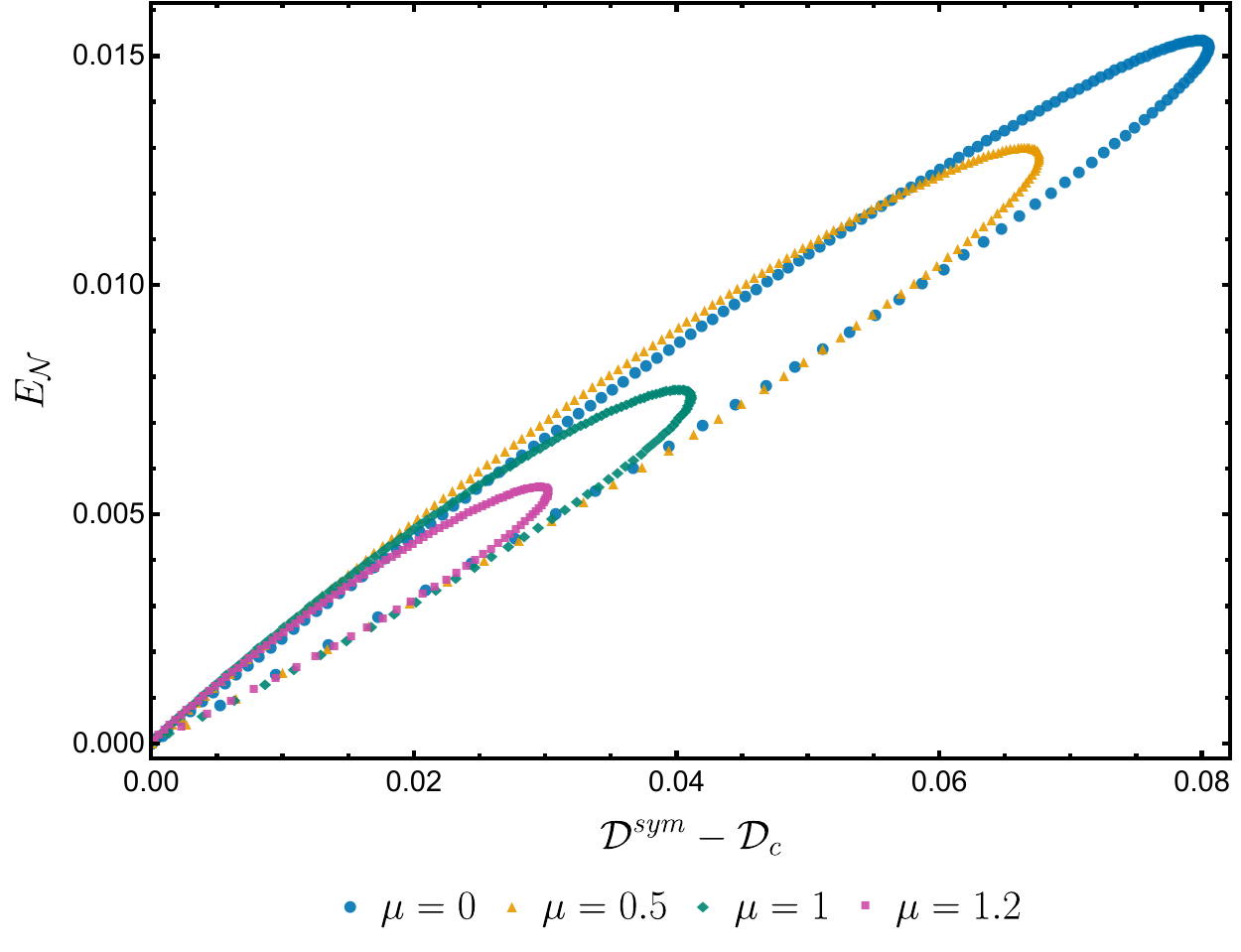}
    \caption{Logarithmic negativity versus $\mathcal{D}^{\mathrm{sym}}-\mathcal{D}_c$ for the ball--shell configuration, where each point corresponds to a \emph{different} value of $d_B$. The rightmost end of each teardrop-shaped curve corresponds to the value $d_B=d_B^{\mathrm{max}}$ for which the entanglement is maximal for the corresponding value of $\mu=m R_A$, while points on the upper and lower branches of each teardrop correspond to values of $d_B$ smaller and larger than $d_B^{\mathrm{max}}$, respectively. We observe that entanglement between $A$ and $B$ occurs only when $\mathcal{D}^{\mathrm{sym}}-\mathcal{D}_c>0$ and both quantities grow monotonically.}
    \label{fig:EN-VS-Ds-Dc}
\end{figure}

To further illustrate this point, in Fig.~\ref{fig:EN-VS-Ds-Dc} we plot the logarithmic negativity $E_{\mathcal{N}}$ versus $\mathcal{D}^{\mathrm{sym}}-\mathcal{D}_c$. Each point in this plot corresponds to a different value of $d_B$. We explicitly see that $\mathcal{D}^{\mathrm{sym}}-\mathcal{D}_c>0$
when $E_{\mathcal{N}}>0$, and both quantities grow monotonically in this example.

In the following subsection, we further explore the relation between these two quantities in the weak-entanglement regime, where an analytic relation can be derived.

\section{Small entanglement regime and Negativity}
This section investigates the relationship between the logarithmic negativity and
\begin{align}
    \mathcal{D}_T \equiv \mathcal{D}^{\mathrm{sym}} - \mathcal{D}_c \, ,
\end{align}
in the regime $|\mathcal{D}_T| \ll 1$—dubbed the \emph{small-entanglement} regime. This regime is particularly relevant when analyzing entanglement between localized modes in a quantum field, as such modes typically exhibit only weak entanglement.

We begin by observing that $\mathcal{D}^{\mathrm{sym}}$ depends linearly on the symplectic invariant $\det J_C$ (see Eq.~\eqref{OverlapFormulaDeterminants}), whereas $\mathcal{D}_c$ is independent of it (see Eq.~\eqref{Dceformula}). On the other hand, the logarithmic negativity between two modes is given by
\begin{equation}\label{LogNegTwomodes}
      E_{\mathcal{N}} = \max\{0,-\log_{2}\tilde{\nu}_{-}\}\, ,
\end{equation}
where $\tilde{\nu}_{-}$ is expressed in Eq.~\eqref{SympEigenValuesTwoModes} in terms of the symplectic invariants $\det J_{AB}$, $\det J_A$, $\det J_B$, and $\det J_C$. By expressing $\det J_C$ in terms of $\mathcal{D}_T$, one can directly determine the dependence of $\tilde{\nu}_{-}$ on $\mathcal{D}_T$. Expanding $-\log_{2}\tilde{\nu}_{-}$ around $\mathcal{D}_T \approx 0$, we obtain
\begin{align}\label{LogNegDTexpansion}
     E_{\mathcal{N}} \simeq \max\{0,\; w(\Delta)\,\mathcal{D}_T 
     + \mathcal{O}(\mathcal{D}_T^{2})\}\, ,
\end{align}
where
\begin{align}
w(\Delta)
   := \frac{1}{\log 2}\,
      \frac{\Delta_A \Delta_B}{\Delta_{AB}(\Delta_A+\Delta_B)}\, ,
\end{align}
with $\Delta_{AB} = \det J_{AB} - 1$ and $\Delta_I = \det J_I - 1$ for $I = A,B$. Since $\Delta_A$, $\Delta_B$, and $\Delta_{AB}$ are positive whenever $A$, $B$, and $AB$ are in mixed states, respectively, the coefficient $w(\Delta)$ is strictly positive in that regime. Therefore, if $\mathcal{D}_T>0$, or equivalently if the modes $A$ and $B$ are entangled, we find
\begin{align}
    E_{\mathcal{N}}
    = w(\Delta)\,\mathcal{D}_T + \mathcal{O}(\mathcal{D}_T^{2}) \, .
    \label{eq:first_order_approximation_negativity}
\end{align}
This expression shows that, in the regime $\mathcal{D}_T \ll 1$, the logarithmic negativity grows linearly with $\mathcal{D}_T$, with a slope determined by the purities of the individual modes $A$, $B$, and the joint system $AB$.

\subsection{Ball–shell example}
To illustrate the usefulness of the relation in Eq.~\eqref{eq:first_order_approximation_negativity}, we apply it to the example discussed in the previous section.

\begin{figure}
    \centering
    \includegraphics[width=0.5\textwidth]{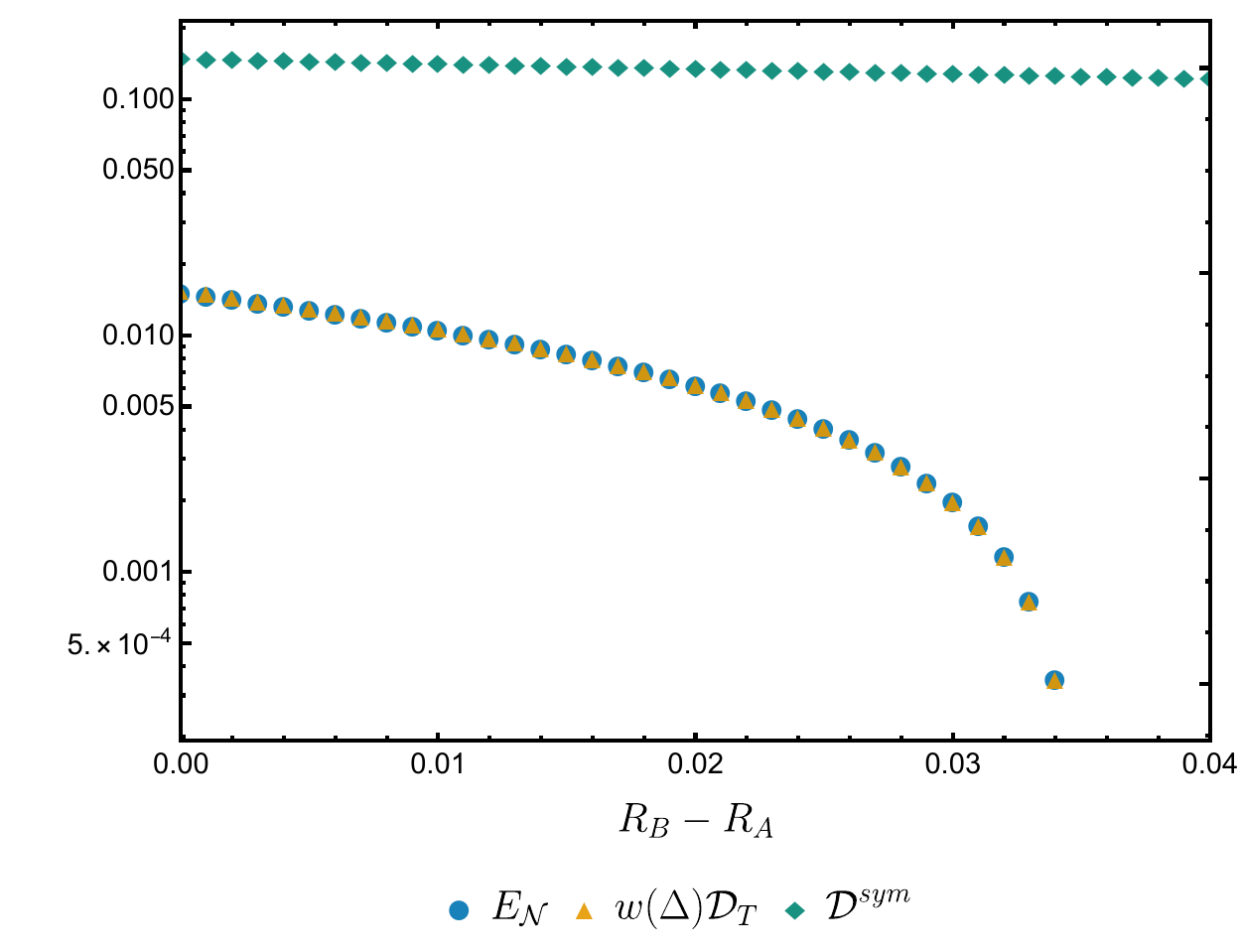}
    \caption{Entanglement related quantities as functions of the separation between the two mode susbystems $R_B-R_A$ in units of $R_A$.  This plot corresponds to $m=0$ and $d_B=0.5 R_A$. The vertical axes is in logarithmic scale. 
    This figure shows that $w(\Delta) \mathcal{D}_T$
    approximates very well the entanglement between the two modes $A$ and $B$ in the entire range of distance in which the two modes are entangled. On the contrary,     $\mathcal{D}^{sym}$ has a much slower decay. 
    }
    \label{fig:EN-LogNegExp-DT-VS-Distance}
\end{figure}

In Fig.~\ref{fig:EN-LogNegExp-DT-VS-Distance}, we plot, for the case of a massless field, both the logarithmic negativity and $w(\Delta)\mathcal{D}_T$ as functions of the radial separation between the two modes, $R_B - R_A$ . We observe that the two quantities are indistinguishable over the entire range of distances for which the two modes are entangled. For completeness, the figure also shows the symmetric overlap $\mathcal{D}^{\mathrm{sym}}$, which exhibits a slow decay with distance and never vanishes entirely---reflecting the presence of long-range correlations in the field theory. The figure clearly demonstrates that the first-order expansion in $ E_{\mathcal{N}}\approx w(\Delta)\mathcal{D}_T$ provides an accurate approximation to the logarithmic negativity for the chosen modes.

\begin{figure}
    \centering
    \includegraphics[width=0.5\textwidth]{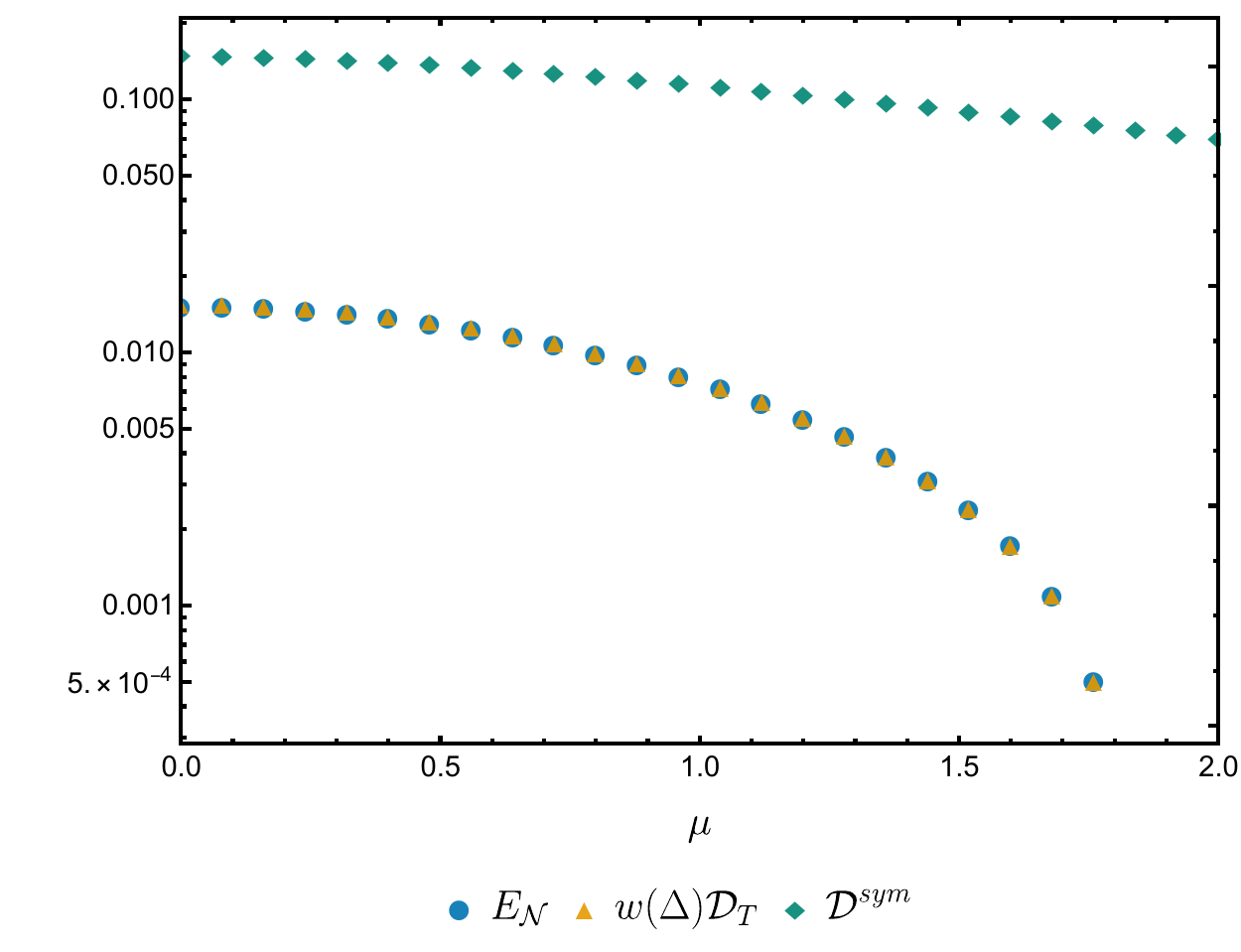}
    \caption{Entanglement related quantities as functions of the dimensionless field mass $\mu=m R_A$.  This plot corresponds to $d_B=0.5 R_A$ and $R_B=R_A$. The vertical axes is in logarithmic scale. The conclusions are the same as in the previous figure.} 
    \label{fig:EN-LogNegExp-DT-VS-mass}
\end{figure}

Finally, in Fig.~\ref{fig:EN-LogNegExp-DT-VS-mass}, we plot the same quantities for various values of the field mass $m$. As the mass increases, all of the quantities decrease, consistent with the fact that the correlation length of the field is of order $1/m$, implying that localized modes at different spatial points become less correlated for larger masses. As shown in the figure, the rapid decay of the logarithmic negativity is extremely well captured by its first-order approximation, $w(\Delta)\mathcal{D}_T$, in Eq.~\eqref{eq:first_order_approximation_negativity}.

These observations establish $w(\Delta)\mathcal{D}_T$ as a useful proxy for entanglement in the weakly correlated regime, which is typically the case for localized modes in quantum field theory. A natural interpretation emerges from this conclusion: the degree of entanglement between localized modes can be understood as a measure of their symmetric overlap, weighted by the factor $w(\Delta)$, which depends on the purity of the two-mode system. On the one hand, increasing the overlap enhances the entanglement between the two modes. On the other hand, $w(\Delta)$ decreases as the reduced state $\hat{\rho}_{AB}$ becomes more mixed, showing that, for the same degree of overlap, modes $A$ and $B$ are less entangled when $\hat{\rho}_{AB}$ is mixed—consistent with the intuition based on entanglement monogamy.

\section{Discussion}

This article introduces a quantifier of correlations between two independent bosonic modes, denoted by $\mathcal{D}^{\mathrm{sym}}$ and dubbed the \emph{symmetric overlap} between the modes. This quantity is invariant under local symplectic transformations within each mode, and it admits a clear intuitive interpretation as the symmetric overlap between each mode and the purification partner of the other. Our $\mathcal{D}^{\mathrm{sym}}$ turns out to be a simple generalization of the correlation quantifier recently proposed in \cite{osawa_entanglement_2025}.

The analysis presented in this article is based on a formulation of continuous-variable systems in terms of a complex version of the classical phase space, shifting the focus from the covariance matrix of Gaussian states to an associated complex structure. Working in this complex framework brings several mathematical and conceptual advantages for analyzing the structure of correlations and entanglement in many-body bosonic systems and quantum fields \cite{agullo_correlation_2025,hackl_bosonic_2021}.

The purpose of this article has been to make precise, for pairs of Gaussian modes, in which sense the purification partners encode the distribution of entanglement with the rest of the system. In the case of quantum fields, this corresponds to how the entanglement is localized in space.

The core result of this article is the derivation of a necessary and sufficient condition for two modes of a Gaussian bosonic system to be entangled, expressed in terms of the symmetric overlap $\mathcal{D}^{\mathrm{sym}}$. The condition reads
\begin{equation}\label{cond}
\mathcal{D}^{\mathrm{sym}} > \mathcal{D}_c \, ,
\end{equation}
where the threshold value $\mathcal{D}_c$ depends on the reduced state of the two modes and, importantly, on the purity of the two-mode system (see Eq.~\eqref{eq:D_c}). 

This entanglement criterion admits an appealing physical interpretation. On the one hand, a larger geometric overlap of each mode with the purification partner of the other mode, quantified by $\mathcal{D}^{\mathrm{sym}}$, favors entanglement between the two modes. As mentioned above, this expresses in mathematical terms the intuitive idea that the spatial support of the partner mode reveals where entanglement is localized within a larger system. For instance, in quantum field theory, if the partner of a given mode is highly localized in a particular spatial region, this indicates that the entanglement resides predominantly in that region.

On the other hand, Eq.~\eqref{cond} shows that a large overlap between two modes $A$ and $B$ does not by itself guarantee entanglement—the overlap must exceed the threshold value $\mathcal{D}_c$. This threshold increases with the mixedness of the two-mode system $(A,B)$, implying that for a fixed geometric overlap the mixedness of the reduced state determines whether the modes are entangled. In this sense, Eq.~\eqref{cond} captures the two essential ingredients that govern entanglement between two modes: their geometric overlap, quantified by $\mathcal{D}^{\mathrm{sym}}$,  and their mixedness, quantified by $\mathcal{D}_c$. Since the mixedness of the subsystem $(A,B)$ grows with its entanglement with other degrees of freedom outside this two-mode subsystem, this condition can be interpreted as a manifestation of entanglement monogamy: entanglement of $(A,B)$ with other modes comes at the expense of entanglement between $A$ and $B$.

We have further shown that, in the low-entanglement regime, the logarithmic negativity (which is a faithful entanglement measure for two Gaussian modes) is proportional to the difference $
\mathcal{D}^{\mathrm{sym}} - \mathcal{D}_c$, 
with a proportionality factor that depends on the mixedness of the two-mode system $(A,B)$. This result further strengthens the interpretation outlined above.

Our results apply both to quantum mechanical systems and to quantum field theories. As an explicit example, we analyzed the entanglement between two modes of a scalar field theory of arbitrary mass $m$, with one mode supported inside a sphere and the other supported in a surrounding spherical shell. This example serves to illustrate and validate the main results of this article.

Taken together, these results offer a geometric perspective on entanglement between two modes within a larger system, complementing other  approaches. Our results also place on quantitative grounds the information encoded in partner modes, clarifying in which precise sense they inform us about the distribution of entanglement within a system.

Finally, the analysis presented here serves to emphasize and promote the effectiveness of complex phase-space methods and complex structures for the study of correlations and entanglement in bosonic Gaussian quantum systems.

\begin{acknowledgments}
I.A. is supported by the NSF grants PHY-2409402 and PHY-2110273, by the RCS program of Louisiana
Boards of Regents through the grant LEQSF(2023-25)-RD-A-
04,  by the Hearne Institute for Theoretical Physics and by Perimeter Institute of Theoretical Physics through the Visitor fellow program. IA is also supported the WOST (WithOut SpaceTime) project (https://withoutspacetime.org), led by the Center for Spacetime and the Quantum (CSTQ), and supported by Grant ID 63683 from the John Templeton Foundation (JTF). The opinions expressed in this work are those of the author(s) and do not necessarily reflect the views of the John Templeton Foundation or those of the Center for Spacetime and the Quantum. E.M-M acknowledges the funding of his NSERC Discovery grant. SNG acknowledges the financial support of the Research Council of Finland through the Finnish Quantum Flagship project (358878, UH). 
K.Y. acknowledges support from the JSPS Overseas Research Fellowship and JSPS KAKENHI Grant No. JP24KJ0085.
 Research at Perimeter Institute is supported
in part by the Government of Canada through the Department of Innovation, Science and Industry Canada
and by the Province of Ontario through the Ministry of
Colleges and Universities.
\end{acknowledgments}

\appendix

\section{Extension to multi-mode Gaussian subsystems}\label{app:multimode_and_mixed}

In the main text, we focused on  correlations and entanglement between two single-mode Gaussian subsystems $A$ and $B$ using the overlaps $\mathcal{D}_{B_p A}$ and $\mathcal{D}_{A_pB}$,  and $\mathcal{D}^{sym}$. Here, we extend this formulation to cases where $A$ and $B$  contain several modes.

Let $\Gamma_X$ and $\Gamma_Y$ be symplectic subspaces of the complexified phase space $\Gamma_{\mathbb{C}}$. Let $\dim \Gamma_X = 2N_X$, and $\{\gamma_I^{(X)},\gamma_I^{(X)*}\}_{I=1}^{N_X}$ be a symplectically orthonormal complex  basis:
\begin{align}
\braket{\gamma_I^{(X)},\gamma_J^{(X)}}&=-\braket{\gamma_I^{(X)*},\gamma_J^{(X)*}}=\delta_{IJ},\\ \braket{\gamma_I^{(X)},\gamma_J^{(X)*}}&=0.
\end{align}
Using such a basis, the symplectic projector onto $\Gamma_X$ can be expressed as
\begin{align}
\Pi_X(\cdot)= \sum_{I=1}^{N_X}\left(\gamma_{I}^{(X)}\braket{\gamma_{I}^{(X)},\cdot}-\gamma_{I}^{(X)*}\braket{\gamma_{I}^{(X)*},\cdot}\right),
\end{align}
which generalizes Eq.~\eqref{eq:projector_in_terms_of_basis_vectors}.

Similarly, let  $\{\gamma_J^{(Y)},\gamma_J^{(Y)*}\}_{J=1}^{N_Y}$ be an orthonormal basis of $\Gamma_Y$,  with $\dim \Gamma_Y = 2N_Y$.

Generalizing Eq.~\eqref{eq:definition_overlap}, we define
\begin{align}
\mathcal{D}_{XY}\coloneqq \sum_{J=1}^{N_Y}\braket{\Pi_X(\gamma_J^{(Y)}),\Pi_X(\gamma_J^{(Y)})},\label{eq:definition_overlap_multi}
\end{align}
which can  be recast as
\begin{align}
\mathcal{D}_{XY}=\sum_{I=1}^{N_X}\sum_{J=1}^{N_Y}\left(|\langle \gamma_J^{(Y)},\gamma_I^{(X)} \rangle|^2-|\langle \gamma_J^{(Y)*},\gamma_I^{(X)} \rangle|^2\right),\label{eq:symmetricity_overlap_multi}
\end{align}
extending Eq.~\eqref{eqOverlap}.

The following arguments justify interpreting $\mathcal{D}_{XY}$ as a measure of the overlap between the symplectic subspaces $\Gamma_X$ and $\Gamma_Y$. First, from Eq.~\eqref{eq:definition_overlap_multi}, $\mathcal{D}_{XY}$ is invariant under symplectic transformations in $\Gamma_X$. Second, from Eq.~\eqref{eq:symmetricity_overlap_multi}, the overlap is symmetric, i.e., $\mathcal{D}_{XY}=\mathcal{D}_{YX}$, and hence also ensures invariance under symplectic transformations in $\Gamma_Y$.

Using the above quantifier of overlap $\mathcal{D}_{XY}$, we now extend the definition of $\mathcal{D}^{sym}$ to the multimode case. Let $A$ and $B$ be independent subsystems, each possibly consisting of multiple modes. We denote by $\Gamma_A\subset \Gamma_{\mathbb{C}}$ and $\Gamma_B\subset \Gamma_{\mathbb{C}}$ the symplectic subspaces characterizing $A$ and $B$, respectively. Assume that the total system is in a pure Gaussian state specified with a complex structure $J$. Then the partner $A_p$ of $A$ is given by~\cite{agullo_correlation_2025} 
\begin{align}
    \Gamma_{A_p}=\Pi_A^\perp (J\Gamma_A)
\end{align}
and similarly
\begin{align}
    \Gamma_{B_p}=\Pi_B^\perp (J\Gamma_B).
\end{align}
We then introduce
\begin{align}
    \mathcal{D}^{sym}\coloneqq \mathcal{D}_{A_pB}+\mathcal{D}_{AB_p}
\end{align}
as a quantifier of overlap between subsystems $A$ and $B$ and their partners. 

As in the single-mode case, the symmetric overlap $\mathcal{D}^{\mathrm{sym}}$ for subsystems $A$ and $B$ satisfies the following properties:
(i) invariance under local symplectic transformations;
(ii) symmetry under exchanging $A$ and $B$;
(iii) $\mathcal{D}^{\mathrm{sym}}=0$ when $A$ and $B$ are uncorrelated;
and (iv) $\mathcal{D}^{\mathrm{sym}}$ depends only on the reduced state $\hat \rho_{AB}$.
Note that Property~(iv) ensures that $\mathcal{D}^{\mathrm{sym}}$ is well defined for mixed Gaussian states by using a Gaussian purification, and that it is independent of the choice of purification, as in the single-mode case.

Property~(i) follows immediately from the symplectic invariance of $\mathcal{D}_{XY}$ together with the partner formula, whereas Property~(ii) is immediate from the definition of $\mathcal{D}^{\mathrm{sym}}$.
Property~(iii) can be proved by repeating the argument used in the single-mode case.
Property~(iv) follows from the following observation: for any $\gamma_{A_p}\in\Gamma_{A_p}$, there exists $\gamma_A\in \Gamma_A$ such that $\gamma_{A_p}=\Pi_A^\perp(J\gamma_A)$.
Therefore,
\begin{align}
    \Pi_{B}(\gamma_{A_p})
    =\Pi_B\left((J\gamma_A)-\Pi_A(J\gamma_A)\right)
    =\Pi_B(J\gamma_A),
\end{align}
where we used the assumption that $A$ and $B$ are symplectically orthogonal.
Consequently,
\begin{align}
    &\braket{\Pi_{B}(\gamma_{A_p}),\Pi_{B}(\gamma_{A_p})}
    \nonumber\\
    &=\sum_{I=1}^{N_{B}}\left(|\braket{\gamma_I^{(B)},J\gamma_A}|^2-|\braket{\gamma_I^{(B)*},J\gamma_A}|^2\right),
\end{align}
where $\{(\gamma_I^{(B)},\gamma_I^{(B)*})\}_{I=1}^{N_{B}}$ is a symplectically orthonormal basis of $\Gamma_B$.
This shows that $\mathcal{D}_{A_pB}$ depends only on the reduced state $\hat{\rho}_{AB}$; hence so does $\mathcal{D}^{\mathrm{sym}}= \mathcal{D}_{A_pB}+\mathcal{D}_{AB_p}$.

\bibliography{references.bib,ref_ky}

\end{document}